\documentclass[12pt,DIV=13]{scrartcl}
\usepackage{dsfont}

\usepackage{lmodern,microtype}
\DeclareOldFontCommand{\bf}{\normalfont\bfseries}{\mathbf}
\DeclareOldFontCommand{\it}{\normalfont\itshape}{\mathit}
\newenvironment{sciabstract}{\begin{abstract}}{\end{abstract}}

\usepackage[colorlinks=true,linkcolor=blue,urlcolor=blue,citecolor=blue,anchorcolor=green,pdfusetitle]{hyperref}
\usepackage{amsmath,amssymb,url,graphicx,mathtools}
\usepackage{tikz}
\input{braket}

\newcommand{\suppref}[2][App]{#1.\ \ref{#2}}

\usepackage{grffile}
\usepackage{longtable,booktabs}

\usepackage[backend=bibtex8,style=alphabetic,doi=false,isbn=false,url=false,maxbibnames=20,sorting=nty,sortcites=false]{biblatex}
\renewbibmacro{in:}{}
\newbibmacro{string+doi}[1]{\iffieldundef{doi}{#1}{\href{https://dx.doi.org/\thefield{doi}}{#1}}}
\DeclareFieldFormat{title}{\usebibmacro{string+doi}{\mkbibemph{#1}}}
\DeclareFieldFormat[article]{title}{\usebibmacro{string+doi}{\mkbibquote{#1}}}
\DeclareFieldFormat[incollection]{title}{\usebibmacro{string+doi}{\mkbibquote{#1}}}

\renewcommand{\eqref}[1]{Eq.\ \ref{#1}}

\newcommand{\cA}{\mathcal{A}}
\newcommand{\cB}{\mathcal{B}}
\newcommand{\cD}{\mathcal{D}}
\newcommand{\cH}{\mathcal{H}}
\newcommand{\cM}{\mathcal{M}}
\newcommand{\cN}{\mathcal{N}}
\newcommand{\cR}{\mathcal{R}}
\newcommand{\cS}{\mathcal{S}}

\newcommand{\Npq}{\cN_{p,q}}
\newcommand{\ox}{\otimes}
\DeclareMathOperator{\poly}{poly}
\DeclareMathOperator{\relu}{ReLU}
\DeclareMathOperator{\sigmoid}{sigm}
\newcommand{\fit}{\mathrm{fit}}
\DeclareMathOperator{\ame}{AME}
\newcommand{\Qone}{Q^{(1)}}
\DeclareMathOperator{\artanh}{artanh}

\newcommand{\op}[1] {\mathbf{#1}}
\newcommand{\field}[1] {\mathds{#1}}
\DeclareMathOperator{\tr}{tr}
\DeclareMathOperator{\id}{id}
\DeclareMathOperator{\PP}{\mathds P}
\DeclareMathOperator{\lin}{span}
\DeclareMathOperator*{\argmax}{argmax}
\DeclareMathOperator{\bin}{bin}

\title{Quantum Codes from Neural Networks} 

\author
{Johannes Bausch,$^{1\ast}$\ \ Felix Leditzky$^{2\dagger}$\\
\\
\normalsize{$^{1}$DAMTP, University of Cambridge, UK,}\\
\normalsize{$^{2}$JILA \& CTQM, University of Colorado Boulder, USA}\\
\\
\normalsize{Email: $^\ast$jkrb2@cam.ac.uk\ \  $^\dagger$felix.leditzky@jila.colorado.edu} 
}

\date{\today}

\begin{document}
\baselineskip17pt

\maketitle
\thispagestyle{empty}
\enlargethispage{1cm}

\begin{sciabstract}
	We examine the usefulness of applying neural networks as a variational state ansatz for many-body quantum systems in the context of quantum information-processing tasks.
	In the neural network state ansatz, the complex amplitude function of a quantum state is computed by a neural network.
	The resulting multipartite entanglement structure captured by this ansatz has proven rich enough to describe the ground states and unitary dynamics of various physical systems of interest.
	In the present paper, we initiate the study of neural network states in quantum information-processing tasks. 
	We demonstrate that neural network states are capable of efficiently representing quantum codes for quantum information transmission and quantum error correction, supplying further evidence for the usefulness of neural network states to describe multipartite entanglement.
	In particular, we show the following main results:
	a) Neural network states yield quantum codes with a high coherent information for two important quantum channels, the generalized amplitude damping channel and the dephrasure channel. These codes outperform all other known codes for these channels, and cannot be found using a direct parametrization of the quantum state.
	b) For the depolarizing channel, the neural network state ansatz reliably finds the best known codes given by repetition codes.
	c) Neural network states can be used to represent absolutely maximally entangled states, a special type of quantum error-correcting codes.
	In all three cases, the neural network state ansatz provides an efficient and versatile means as a variational parametrization of these highly entangled states.
\end{sciabstract}

\newpage
\tableofcontents

\section{Introduction}
The exponential growth of the Hilbert space dimension in the number of particles is both a blessing and curse for quantum science:
On the one hand, it is crucial to the widely-believed computational advantage of quantum computers over classical ones, but on the other hand it renders many questions about properties of many-body systems intractable.
Yet we know that the ``physical'' corner of this Hilbert space has to be small: local Hamiltonians with highly-entangled ground states only require a polynomial number of parameters to describe, as do quantum circuits of polynomial depth.

This fact motivates the use of variational representations of quantum states to solve a large class of problems.
At the heart of any variational ansatz is the idea to preserve as much information about the quantum state as possible, while discarding irrelevant features.
Quantum mechanical properties of a state are fundamentally dictated by its entanglement, which captures quantum correlations between its subsystems.

For instance, correlation length in many-body spin systems is tightly linked to the existence of a spectral gap \cite{Hastings2007,Gosset2016}.
For gapped one-dimensional systems (which follow an entanglement entropy area law), one can use matrix product states (MPS) with polynomial bond dimension to efficiently represent ground states \cite{FNW92,Landau2013,Arad2013}.
The MPS ansatz has further proven useful e.g.\ in the study of cricital systems \cite{Pirvu2012} or in the continuum limit \cite{Cuevas2017}.
Other tensor network states include MERA and higher-dimensional variants such as PEPS---applied e.g.\ in the context of renormalization \cite{Vid07,VC04}, and proven similarly successful as part of numerical techniques \cite{Orus2014}.

A relatively recent development is the use of neural network states as a variational ansatz, where the network is used as a function to calculate the state amplitudes \cite{CT17}.
There are many possible neural network architectures to choose from:
one proposed model is to use restricted Boltzmann machines (RBMs) to represent e.g.\ the ground states and unitary dynamics of a transverse-field Ising model and the antiferromagnetic Heisenberg model \cite{CT17}, volume-law entanglement and the ground state of even long-range Hamiltonians \cite{Deng2017}, as well as ground states of various stabilizer Hamiltonians, including the surface code \cite{Jia2018}.
While there exist local Hamiltonians that cannot be represented efficiently with shallow RBM architectures, it has been shown that deep RBM networks can in fact represent {\it most} physical states, which includes those that can be created by poly-depth quantum circuits, or ground states of local Hamiltonians with a $1/\poly$ spectral gap \cite{Gao2017}.

Apart from describing the physics of many-body systems, entanglement also plays a crucial role in information-processing tasks: teleportation \cite{BBC+93}, superdense coding \cite{BW92}, and entanglement-assisted classical \cite{BSST99} and quantum \cite{DHW04} communication all build on \emph{bipartite} entanglement as a resource.
In contrast, for certain tasks such as quantum information transmission through many uses of a quantum channel, or the encoding of quantum information in quantum error correction codes, the crucial property is \emph{multipartite entanglement}, which encapsulates correlations among all the constituents of the system simultaneously \cite{Hastings2007}.

\subsection{Main Results}
We demonstrate that neural network states with only polynomially many parameters in the system size (which, in this context, we call \emph{efficient}) are capable of representing quantum codes for quantum information transmission and quantum error correction.
In particular, we show the following:
\begin{itemize}
	\item The neural network state ansatz finds new quantum codes with a high coherent information (CI) that outperform all previously known codes for two channel models, the generalized amplitude damping channel and the dephrasure channel.
	For the generalized amplitude damping channel, the new codes also increase the threshold of the channel, i.e., the boundary of the interval in the parameter space with positive quantum capacity.
	For both channels, the new codes cannot be found with `traditional' numerical methods, i.e., a direct parametrization of the complex amplitudes of the quantum state.
	\item For the depolarizing channel, neural network states can efficiently represent the best known codes.
	We carry out a detailed comparison of different network architectures, showing that FF networks converge faster than RBMs with comparable parameter counts in almost all tested cases.
	Furthermore, we constructively prove that the best known codes (repetition codes, and products thereof) can be obtained efficiently with both an RBM and a FF architecture.
	\item Neural network states can be used to parametrize so-called ``absolutely maximally entangled'' (AME) states.
		These AME states, defined on $n$ systems of local dimension $d$ each, are examples of quantum error-correcting codes with the property that they are completely mixed after tracing out at least half of the systems. 
		Besides their quantum error correction capabilities, AME states are useful in multi-user information-theoretic tasks such as open-destination teleporation, secret sharing or entanglement swapping that require maximal entanglement across different choices of bipartitions \cite{HC13,HCLRL12}.
\end{itemize}
The properties of both quantum codes with high coherent information and AME states are the result of the non-trivial multipartite entanglement present in these states.
The main finding of this paper is that for both high-CI states and AME states, a neural network state ansatz is able to faithfully represent this multipartite entanglement, which we demonstrate empirically for small problem instances.
We furthermore provide numerical evidence that the variational ansatz vastly outperforms a full state parametrization for the respective learning tasks.

\subsection{Structure of this Paper}
This paper is structured as follows.
In Sec.~\ref{sec:capacity} we introduce the quantum capacity of a channel and state the corresponding coding theorem which expresses the quantum capacity as a regularized formula in terms of an entropic quantity called the coherent information.
We then discuss how lower bounds on the quantum capacity can be obtained by solving an entropic optimization problem.
In Sec.~\ref{sec:NN-states} we review neural network states based on restricted Boltzmann machines and feed-forward nets.
We then present our main results about representing quantum codes with neural network states.
In Sec.~\ref{sec:new-codes} we discuss the generalized amplitude damping channel and the dephrasure channel.
We show that the neural network state ansatz finds new quantum codes providing the strongest lower bounds to date on the quantum capacities of these channels.
Moreover, we demonstrate that these new codes are not found using a ``direct'' parametrization of quantum states.
We then show in Sec.~\ref{sec:depolarizing} for the depolarizing channel how tensor products of repetition codes---i.e.\ the known optimal codes for $k\leq 9$ uses of this channel---can be efficiently represented using FF and RBM networks, and comment on the trainability of our chosen network architectures.
Finally, in Sec.~\ref{sec:ame-main} we demonstrate how known examples of AME states can be efficiently represented using neural networks, and we comment on the trainability of the network architectures that we used.
We conclude in Sec.~\ref{sec:discussion} with a discussion of our results and open problems.

In the appendices, we give more details about certain aspects of the paper.
In \suppref{sec:gadc-codes} and \suppref{sec:dephrasure-codes} we state explicit formulas for the coherent information of weighted repetition codes for the generalized amplitude damping channel and dephrasure channel, respectively, which serve as benchmarks for our quantum codes from neural networks.
We also supply additional data obtained in our numerical investigations.
In \suppref{sec:depolarizing-codes} we give an overview of the best known codes for the depolarizing channel, and provide an analytical construction of these codes for neural networks with various architectures.
In \suppref{sec:ame} we provide some background information on absolutely maximally entangled states, and prove a useful bound on a trace distance parameter indicating how close a state is to being absolutely maximally entangled.
In \suppref{sec:encodings} we discuss possible encodings of $d$-ary input strings to neural networks.
In \suppref{sec:activation-f} we comment on the role of activation functions for quantum codes; furthermore, we propose a novel NN Schmidt decomposition ansatz, which we benchmark against a full NN parametrization for the depolarizing channel.
In  \suppref{sec:numerics} we give a high-level explanation of the global derivative-free numerical optimization techniques used in our paper.
Finally, we provide additional numerical data for some of our results in \suppref{sec:extra-numerics}.

We encourage researchers to adopt our methods by providing full access to our code (in C++ and MATLAB) that was used to obtain the numerical results of this paper.
These code files can be found in the ``Ancillary files'' section of the arXiv post of this paper \cite{anc_files}.
In MATLAB, we made use of the MATLAB Global Optimization Toolbox, as well as quantinf \cite{Cub} and QETLAB \cite{qetlab}.
In C++, we made use of NLopt \cite{nlopt} with the CCSA  algorithm \cite{Svan02} as well as PAGMO \cite{pagmo}.

\begin{figure}
	\begin{center}
		\begin{tikzpicture}[
		baseline=0,
		every node/.style={
			circle, draw=black, minimum size=.6cm, inner sep=0pt
		}
		]
		\def\dx{1.2}
		\def\dy{2}
		\foreach \x/\X in {1/2}
		\foreach \a in {1,2,...,5}
		\foreach \b in {1,2,...,5}
		\draw (\dx*\a,\dy*\x) -- (\dx*\b,\dy*\X);

		\foreach \y in {1,2,...,5}
		\draw (\dx*\y,\dy) node[fill=black!20] {$i_\y$};
		\foreach \x/\l in {2/1}
		\foreach \y in {1,2,...,5} {
			\draw (\dx*\y,\dy*\x) node[fill=white] {$h_\y$};
		}
		\end{tikzpicture}
		\hfill
		\begin{tikzpicture}[
		every node/.style={
			circle, draw=black, minimum size=.6cm, inner sep=0pt
		}
		]
		\def\dx{2}
		\def\dy{1.2}
		\foreach \x/\X in {1/2,2/3,3/4}
		\foreach \a in {1,2,...,5}
		\foreach \b in {1,2,...,5}
		\draw (\dx*\x,\dy*\a) -- (\dx*\X,\dy*\b);
		
		\foreach \a in {1,2,...,5} {
			\draw[] (\dx*4,\dy*\a) -- (\dx*5,\dy*3.5);
			\draw[] (\dx*4,\dy*\a) -- (\dx*5,\dy*2.5);
		}
		
		\foreach \y in {1,2,...,5}
		\draw (\dx*1,\dy*\y) node[fill=black!20] {$i_\y$};
		\foreach \x/\l in {2/1,3/2,4/3}
		\foreach \y in {1,2,...,5} {
			\draw (\dx*\x,\dy*\y) node[fill=white] {$f_\l$};
		}
		\draw (\dx*5,\dy*3.5) node[fill=black!20] {$o_1$};
		\draw (\dx*5,\dy*2.5) node[fill=black!20] {$o_2$};
		\end{tikzpicture}
	\end{center}
	\caption{Left: Restricted Boltzmann machine (RBM) with five input nodes and five hidden nodes. Right: Feed forward neural network with five input nodes, two (real-valued) output nodes, and three fully-connected hidden layers of size five each.
		Each line represents one real value being propagated forward from node to node; the $f_i$ are non-linear activation functions (e.g.\ sigmoid, ReLU, $\cos$, see Sec.\ \ref{sec:activation-f} for a discussion) applied to an affine transformation of the node inputs (see \eqref{eq:FF-architecture}).}
	\label{fig:networks}
\end{figure}
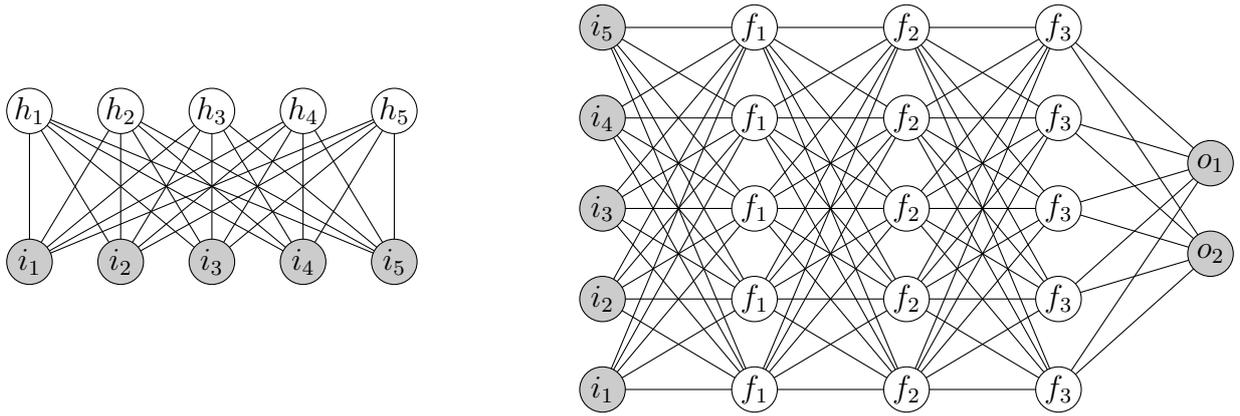

\section{The Quantum Capacity of a Quantum Channel}\label{sec:capacity}
A point-to-point communication link between quantum systems can be modeled by a \emph{quantum channel}.
For quantum systems $A$ and $B$ with underlying (finite-dimensional) Hilbert spaces $\cH_A$ and $\cH_B$, respectively, a quantum channel $\cN\colon A\to B$ is a linear, completely positive, trace-preserving map between the algebras of linear operators $\cB(\cH_A)$ and $\cB(\cH_B)$.
A quantum state $\rho_A$ on $A$ is a linear positive semidefinite operator with unit trace. 
A quantum state $\psi_A$ with rank $1$ is called pure, and can be identified with a normalized vector $|\psi\rangle_A\in\cH_A$ such that $\psi_A = |\psi\rangle\langle \psi|_A$.

The communication capabilities of a quantum channel are characterized by various \emph{capacities}, depending on what kind of information one attempts to transmit faithfully through the channel.
The \emph{quantum capacity} $Q(\cN)$ of a quantum channel $\cN\colon A\to B$ characterizes the optimal rate of faithful quantum information transmission through the channel.
$Q(\cN)$ can be defined in terms of the operational task of \emph{entanglement generation} as follows.

Suppose Alice, the sender, prepares a pure state $\psi_{RA^n}$ in her laboratory and sends the $A^n$-part to Bob through $n$ independent uses of the quantum channel $\cN$.\footnote{That is, Alice uses the $A^n$-part of $\psi_{RA^n}$ as the input to the channel $\cN^{\ox n}$.}
Upon receiving the quantum systems from Alice, Bob applies some decoding operation $\cD_n\colon B^n\to R'$ to the output, yielding the final state $\sigma_{RR'} = (\id_R \ox \cD_n\circ \cN^{\ox n})(\psi_{RA^n})$.
The goal for Alice and Bob is to obtain a final state $\sigma_{RR'}$ that is close (in a suitable distance measure) to a maximally entangled state $|\Phi^{M_n}\rangle_{RR'}$ of Schmidt rank $M_n$, i.e., $|\Phi^{M_n}\rangle_{RR'} = M_n^{-1/2}\sum_{i=1}^{M_n} |i\rangle_R\ox |i\rangle_{R'}$ for some basis $\{ |i\rangle\}_i$.
If there is an entanglement generation protocol such that $\sigma_{RR'}$ converges to $\Phi^{M_n}_{RR'}$ with respect to the chosen distance as $n\to\infty$, then $\lim_{n\to\infty}\frac{1}{n}\log M_n$ is called an \emph{achievable rate}.
The quantum capacity $Q(\cN)$ is defined as the supremum over all achievable rates.

$Q(\cN)$ can be expressed in terms of a \textit{regularized formula} as \cite{Llo97,Sho02,Dev05}
\begin{equation}
Q(\cN)=\lim_{n\to\infty} \frac1n \Qone(\cN^{\otimes n}) = \sup_{n\in\field N} \frac1n \Qone(\cN^{\otimes n}),\label{eq:quantum-capacity}
\end{equation}
where the \emph{channel coherent information} $\Qone(\cN)$ is defined as
\begin{equation}\label{eq:coherent-information}
\Qone(\cN) = \max_{|\psi\rangle_{RA}} \Qone(\psi_{RA},\cN)=\max_{\psi_{RA}} S(\cN(\psi_A)) - S((\id_R\ox\cN)(\psi_{RA} )),
\end{equation}
with the von Neumann entropy $S(\rho):=-\tr(\rho\log(\rho))$.

Formula \ref{eq:quantum-capacity} for the quantum capacity involves the evaluation of the channel coherent information $Q^{(1)}(\cdot)$ over an (in principle) unbounded number of channel copies.
If the channel coherent information is \emph{weakly additive}, $Q^{(1)}(\cN^{\ox n})\leq n Q^{(1)}(\cN)$, then the regularization disappears and \eqref{eq:quantum-capacity} becomes $Q(\cN) = Q^{(1)}(\cN)$.
Weak additivity of the channel coherent information is only known to hold for certain classes of channels such as degradable channels \cite{DS05}.
Moreover, there are examples of quantum channels for which the channel coherent information is strictly \emph{superadditive}, $Q^{(1)}(\cN^{\ox n}) > n Q^{(1)}(\cN)$ for some $n$, rendering the regularization over $n$ in the quantum capacity formula \ref{eq:quantum-capacity} necessary in general \cite{DSS98}. 
However, for so-called \emph{low-noise} channels that are close in diamond norm to a noiseless channel, the effect of superadditivity of coherent information cannot be too large, and the single-letter coherent information is essentially the right answer \cite{LLS17,SSWR15}.
In this paper, we are interested in the \textit{high-noise} regime where superadditivity of channel coherent information typically occurs.

An important part of the quantum capacity theorem in \eqref{eq:quantum-capacity} is the fact that the channel coherent information is an \emph{achievable rate} \cite{Llo97,Sho02,Dev05}:
\begin{align}
Q(\cN) \geq Q^{(1)}(\cN).\label{eq:achievability}
\end{align}
Using block codes, this can be generalized to $Q(\cN) \geq \frac{1}{n} Q^{(1)}(\cN^{\ox n})$ for all $n\in\mathbb{N}$.
The rough proof idea of \eqref{eq:achievability} is the following: Assume that $|\psi\rangle_{RA}$ is a pure state with strictly positive coherent information, $Q^{(1)}(\psi,\cN)>0$.
Once Alice and Bob share $k$ copies of the state $\sigma_{RB}=(\id_R\ox \cN)(\psi_{RA})$ (which they can achieve by Alice sending the $A^k$ part of the state $\psi_{RA}^{\ox k}$ to Bob through $\cN^{\ox k}$) for a sufficiently large $k$, there is a protocol defined in terms of the typical subspaces of $\sigma_{RB}^{\ox k}$ that allows Alice and Bob to generate entanglement between them at a rate of $r-\delta$ for arbitrarily small $\delta\in(0,r)$, where $r$ is equal to the coherent information of the state $\sigma$, that is, $r = I(R\rangle B)_\sigma = Q^{(1)}(\psi,\cN)$ \cite{Dev05,HHWY08}.
Here, $I(R\rangle B)_\sigma = S(\sigma_B) - S(\sigma_{RB})$ is the coherent information of the bipartite state $\sigma_{RB}$.

In this operational picture, we can think of $\psi_{RA}$ as the \emph{inner code}, whereas the (1-LOCC assisted) distillation protocol manipulating $\sigma_{RB}^{\ox k}$ is the \emph{outer code}.
The rate at which the full protocol generates entanglement is solely determined by the (strictly positive) coherent information of the inner code $\psi_{RA}$.
Hence, in this paper we refer to the inner code $\psi_{RA}$ simply as a \emph{quantum code}.
The main objective of this paper is to find quantum codes $|\psi\rangle_{RA^n}$ that achieve high coherent information $\frac{1}{n}\Qone(\psi_{RA^n},\cN^{\ox n})>0$.
To find such quantum codes, we use the neural network state ansatz introduced in \cite{CT17}.
In the next section, we review different variants of this ansatz. 

\section{Neural Network States}\label{sec:NN-states}
For simplicity we consider in the following a system consisting of $n$ qubits, that is, a collection of $n$ $2$-dimensional quantum systems each described by a Hilbert space isomorphic to $\mathbb{C}^2$.
The state space of the $n$ qubits is described by the tensor space $(\mathbb{C}^2)^{\ox n}$ with the ``computational basis'' $\{ |0\rangle,|1\rangle\}^{\ox n}$, and a general pure normalized quantum state $|\psi\rangle\in\cH^{\otimes n}$ can be written as
\begin{align}
|\psi\rangle  = \frac{1}{C}\sum_{i_1,\dots,i_n=0,1} \psi(i_1,\dots,i_n) |i_1\rangle \ox \dots \ox |i_n\rangle = \frac{1}{C}\sum_{i^n\in \{ 0,1\}^n} \psi(i^n) |i^n\rangle. \label{eq:many-body-state}
\end{align}
Here, $C$ is a normalization constant ensuring $\langle \psi|\psi\rangle = 1$, the set of binary strings of length $n$ is denoted by $\{ 0,1\}^n$, and for a string $i^n=(i_1,\dots,i_n)\in\{ 0,1\}^n$ we define $|i^n\rangle \coloneqq |i_1\rangle \ox \dots \ox |i_n\rangle$.
Evidently, a full description of the quantum state $|\psi\rangle$ consists of a list of the $2^n$ complex amplitudes $\psi(i^n)$, corresponding to $2\cdot2^n-1$ real degrees of freedom.

For a neural network state $\psi$, the amplitude function $\psi(i^n)$ in \eqref{eq:many-body-state} is computed from the input string $i^n$ using a neural network.
There are different network architectures that can be used, and we describe a few common choices in the following subsections.

\subsection{Restricted Boltzmann States}\label{sec:1-rbm-states}
The first architecture---and one of the most well-studied ones, see e.g.\ \cite{Glasser2018} for an excellent review---are restricted Boltzmann machines (RBM).
They have proven particularly fruitful as a variational ansatz for representing various ground states of local Hamiltonians \cite{CT17}, notably surpassing fidelity as compared to other neural network architectures in some cases.

A Boltzmann machine has visible and hidden nodes (see Fig.\ \ref{fig:networks}).
A set of complex variables is assigned to each node; we denote the visible units with $i_1,\ldots,i_n$, and the hidden units with $h_1,\ldots,h_m$.
Each link between nodes corresponds to an Ising-type coupling, which defines an energy function (which one can think of as a Hamiltonian)
\begin{equation}\label{eq:rbm-energy}
\op H_\text{RBM} = \sum_{l=1}^n a_l i_l + \sum_{l=1}^m b_l h_l + \sum_{k<l} W_{kl} i_k h_l.
\end{equation}
The two vectors $a\in\field C^n$ and $b\in\field C^m$ define a bias over the visible and hidden nodes, respectively, while the matrix $W\in\field C^{m\times n}$ defines the coupling between the two layers.
The energy of the system allows us to define a complex probability distribution over the vectors $i$ and $h$ via $\PP(i,h):=\exp(-\op H(i,h))/Z$ with partition function $Z=\sum_{i,k}\PP(i,h)$.

To extract a weight $\psi(i^n)$ used to assemble a state via \eqref{eq:many-body-state}, we simply trace out the hidden nodes of the RBM, which yields a marginal probability distribution over the input nodes.
We obtain
\begin{equation}\label{eq:rbm-state}
\ket{\psi_\text{RBM}} = \sum_{i^n\in\{0,1\}^n}\sum_{h^n\in\{0,1\}^n}\frac{\exp(-\op H(i^n,h^n))}{Z}\ket{i^n}.
\end{equation}
If we take all parameters $a, b$ and $W$ to be real-valued, the resulting state will only have real non-negative weights.
In order to retain full generality in the RBM ansatz, the network weights are typically chosen to be complex \cite{CT17}.

\subsection{Deep Boltzmann States}\label{sec:1-dbm-states}
\begin{figure}
	\begin{center}
		\begin{tikzpicture}[
		baseline=0,
		every node/.style={
			circle, draw=black, minimum size=.6cm, inner sep=0pt
		}
		]
		\def\dx{1.2}
		\def\dy{2}
		\foreach \x/\X in {1/2} {
			\foreach \a in {1,2,...,5}
			\foreach \b in {1,2,...,5}
			\draw (\dx*\a,\dy*\x) -- (\dx*\b,\dy*\X);
			\foreach \a in {1,2,...,5}
			\foreach \b in {1,2,...,\a} {
				\draw (\dx*\a,\dy*\x) to[out=-90,in=-90] (\dx*\b,\dy*\x);
				\draw (\dx*\a,\dy*\X) to[out=90,in=90] (\dx*\b,\dy*\X);
			}
			
		};

		\foreach \y in {1,2,...,5}
		\draw (\dx*\y,\dy) node[fill=black!20] {$i_\y$};
		\foreach \x/\l in {2/1}
		\foreach \y in {1,2,...,5} {
			\draw (\dx*\y,\dy*\x) node[fill=white] {$h_\y$};
		}
		\end{tikzpicture}	\end{center}
	\caption{Deep Boltzmann machine (DBM) with five input nodes and five hidden nodes.
		The architecture resembles that of an RBM (see Fig.\ \ref{fig:networks}), but where the nodes within each layer are cross-linked.
		\cite{Gao2017} showed that the model with connections within a layer is equivalent to one with more than two inter-connected layers but no connections within each layer.}
	\label{fig:dbm}
\end{figure}
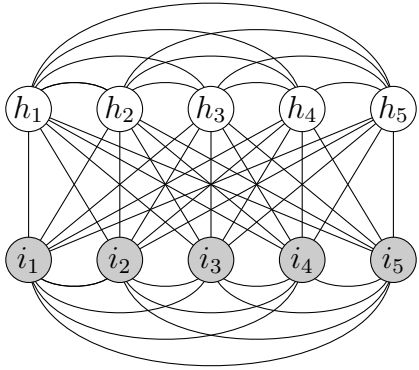
While RBM states struggle to represent e.g.\ ground states for local Hamiltonians with even mildly-decaying spectral gap,
adding links between the nodes within each layer yields a model with vastly greater representative power \cite{Gao2017,Glasser2018}---deep Boltzmann machines (DBMs, see Fig.\ \ref{fig:dbm}).

In analogy to \eqref{eq:rbm-energy}, we can define an energy function for a DBM by introducing additional coupling matrices $D\in\field C^{m\times m}$ and $C\in\field C^{n\times n}$ for the hidden and visible nodes, respectively.
This yields an overall Hamiltonian
\begin{equation}\label{eq:dbm-energy}
\op H_\text{DBM} = \op H_\text{RBM} + \sum_{k<l} C_{kl} i_k i_l + \sum_{k<l} D_{kl} h_k h_l.
\end{equation}
The way one obtains a state from a DBM follows the same method as for an RBM.

\subsection{Feed-Forward Network States}\label{sec:1-ff-states}
The third architecture is obtained by using the most prominent neural network model to date, feed-forward nets, to represent quantum states.
This has proven successful in a number of cases \cite{CL18,Saito2017}.

A feed-forward network consists of a visible layer $v=i^n$ with input nodes $i_1,\dots,i_n$, a fixed number $H$ of hidden layers $h^{(j)}$ of width $M_j$, and an output layer $o$ with two output nodes $o_1$ and $o_2$ (see Fig.\ \ref{fig:networks}).
Each hidden neuron $h^{(j)}_{k}$ for $j\in [H]$ and $k\in [M_j]$ is assigned a \emph{bias} $b^{(j)}_{k}$.
Here, we use the notation $[n]\coloneqq \lbrace 1, \dots, n\rbrace$ for $n\in\mathbb{N}$.
The interactions between two hidden layers $h^{(j-1)}$ and $h^{(j)}$ are mediated by \emph{weight matrices} $(W^{(j)}_{kl})_{kl}$ where $k\in[M_j]$ and $l\in[M_{j-1}]$.
The weight matrix $W^{(1)}$ mediates between the visible layer and the first hidden layer, and the weight matrix $W^{(H+1)}$ mediates between the last hidden layer $h^{(H)}$ and the output layer $o$ with bias $b^{{(H+1)}}$.
In each hidden layer $h^{(j)}$ the state of the neurons is processed with a non-linear activation function $f_j$.
In the following, we interpret the visible layer $v$, the hidden layers $h^{(j)}$, and the output layer $o$ as column vectors, and functions are evaluated component-wise.
Given the input $v=i^n$, the amplitude function $\psi(i^n)$ is computed as follows:
\begin{align}
\begin{aligned}
h^{(1)} &= f_1\left(W^{(1)}v + b^{(1)}\right)\\
h^{(j)} &= f_j\left(W^{(j)} h^{(j-1)} + b^{(j)}\right) \qquad\text{for $j=2,\dots,H$}\\
o &= W^{(H+1)} h^{(H)} + b^{(H+1)}\\
\text{Cartesian: }\quad \psi(i^n) &= o_1 + io_2\\
\text{Polar: }\quad \psi(i^n) &= \exp(o_1 + io_2).
\end{aligned}
\label{eq:FF-architecture}
\end{align}

A network architecture is specified by the data $(H,\{ M_j,f_j\}_{j\in [H]})$.
Common choices for the activation functions are the sigmoid function $\sigma(x)\coloneqq (1+\exp(-x))^{-1}$, the hyperbolic tangent $\tanh$, or the rectified linear unit $\relu(x) = \max\{ 0, x\}$, which are depicted in Fig.~\ref{fig:activation-f}.
From a theoretical point of view these choices are all equivalent, since feed-forward networks as described above are \emph{universal}: With a single hidden layer, they can approximate any given function to arbitrary precision provided the activation function is non-constant and the number of hidden neurons is sufficiently large \cite{Kol61,Hor91}.
However, in practice the choice of activation functions has to be tailored to the problem at hand to achieve good numerical results.
In \suppref{sec:activation-f}, we elaborate on the heuristics of choosing activation functions for neural network states; of particular interest in this context is that periodic activation functions such as cosine seem to be able to capture more of the structure of various quantum states \cite{CL18}.
We prove analytically in \suppref{sec:analytical-proofs} that periodic activation functions are also beneficial in representing good quantum codes.


\section{New Quantum Codes Using a Neural Network State Ansatz}\label{sec:new-codes}

\subsection{Generalized Amplitude Damping Channel}\label{sec:gadc}

The first quantum channel for which we investigate the neural network state ansatz is the \textit{generalized amplitude damping channel} (GADC) $\cA_{\gamma,N}$.
It is defined in terms of two parameters $\gamma,N\in[0,1]$ and acts on a qubit state $\rho$ as $\cA_{\gamma,N} = \sum\nolimits_{i=1}^4 A_i\rho A_i^\dagger$, where
\begin{align}
\begin{aligned}
 A_1 &= \sqrt{1-N}(|0\rangle\langle 0| + \sqrt{1-\gamma}|1\rangle\langle 1|)\\
 A_2 &= \sqrt{\gamma(1-N)} |0\rangle\langle 1|\\
 A_3 &= \sqrt{N} (\sqrt{1-\gamma} |0\rangle\langle 0| + |1\rangle\langle 1|) \\
 A_4 &= \sqrt{\gamma N} |1\rangle \langle 0|.
\end{aligned}\label{eq:gadc}
\end{align}
The GADC models the dynamics of a qubit in contact with a thermal bath at temperature $N$ and transition probability $\gamma$ between the ground state $|0\rangle$ and the excited state $|1\rangle$.
This quantum channel is a realistic noise model in various physical processes such as relaxation processes of spin systems, superconducting quantum computers, and loss processes in linear optical systems \cite{myatt2000decoherence,turchette2000,chirolli2008,zou2017}.
Furthermore, for $N=0$ the GADC reduces to the well-known amplitude damping channel modeling energy dissipation of a qubit.

\begin{figure}
	\centering
	\includegraphics[height=\dimexpr\textheight-130.35004pt\relax]{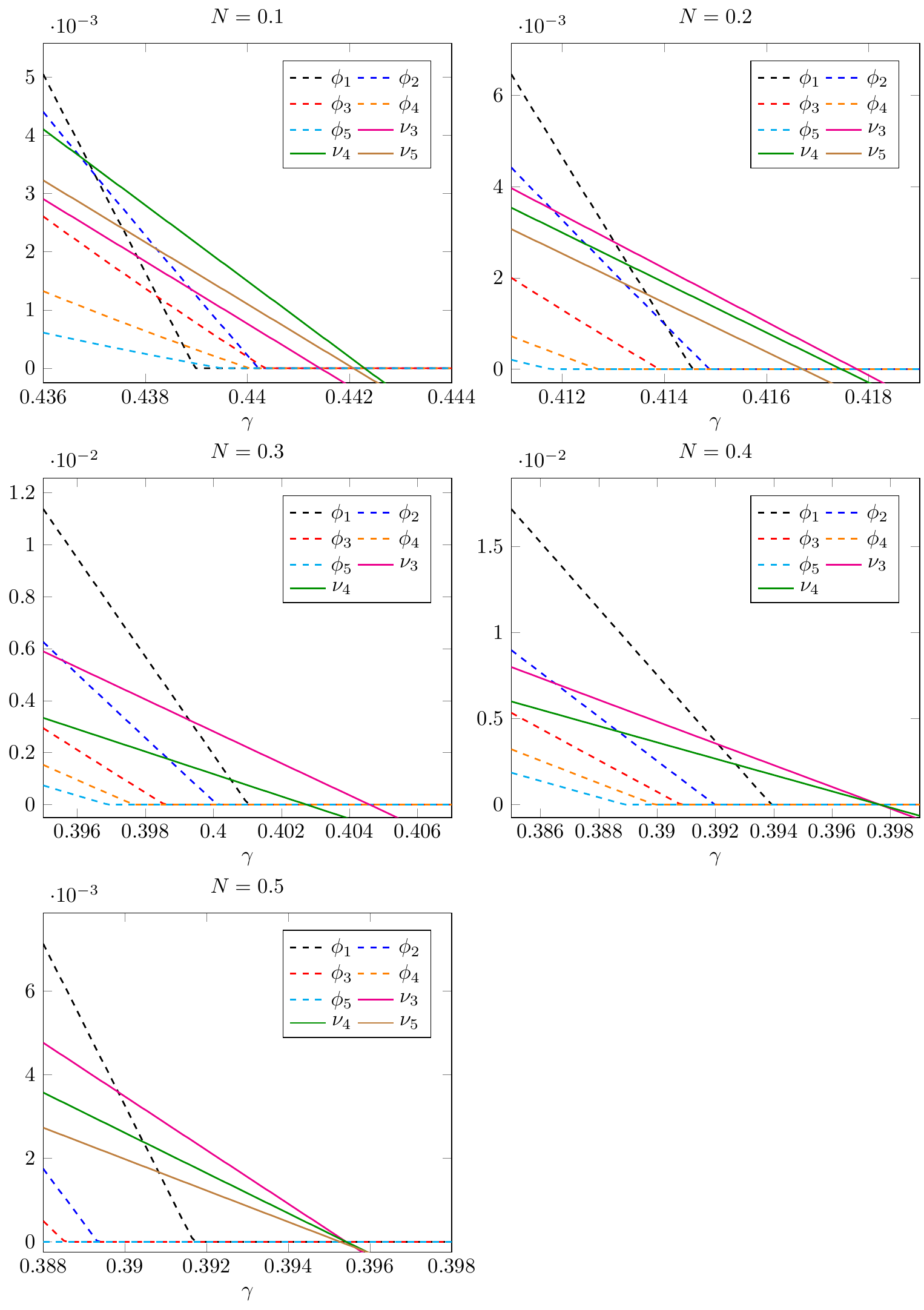}
	\caption{Overview of quantum codes for the generalized amplitude damping channel $\cA_{\gamma,N}$ comparing the neural network codes $\nu_k$ (solid lines) for $k=3,4,5$ to the weighted repetition codes $\phi_k$ (dashed lines) for $1\leq k\leq 5$ defined in \eqref{eq:weighted-repetition-code}. 
	For each $N\in\lbrace 0.1,0.2,0.3,0.4,0.5\rbrace$, we plot the interval of $\gamma$ in which superadditivity occurs. 
	In each case, the neural network codes $\nu_k$ increase the threshold for $\cA_{\gamma,N}$ over the weighted repetition codes. 
	For $N=0.3,0.4$ and $k=5$ copies of $\cA_{\gamma,N}$ the neural network ansatz only found trivial product codes, which are not shown.
	The neural network codes $\nu_k$ are listed in Tabs.~\ref{tab:gadc-codes-N01}, \ref{tab:gadc-codes-N02-app}, \ref{tab:gadc-codes-N03-app}, \ref{tab:gadc-codes-N04-app}, \ref{tab:gadc-codes-N05-app} for $N=0.1,0.2,0.3,0.4,0.5$, respectively.
	}
	\label{fig:gadc-codes}
\end{figure}

While the quantum capacity of the amplitude damping channel $\cA_{\gamma,0}$ is equal to its (additive) single-letter coherent information for all $\gamma\in[0,1]$ and can be computed efficiently \cite{GF05}, the quantum capacity of the more general noise model $\cA_{\gamma,N}$ with $N\in(0,1)$ is unknown.
Various upper bounds on $Q(\cA_{\gamma,N})$ have been computed in the recent work \cite{KSW19}, but so far achievable rates (i.e., lower bounds on $Q(\cA_{\gamma,N})$) improving upon the single-letter coherent information $\Qone(\cA_{\gamma,N})$ have not been studied extensively.
We prove in this section that for $N\in(0,1)$ and particular intervals of $\gamma$ the channel coherent information $\Qone(\cA_{\gamma,N})$ of the GADC is superadditive.
As shown in the discussion below and in Fig.~\ref{fig:gadc-codes}, superadditivity is achieved by, e.g., weighted repetition codes 
\begin{align}
|\phi_k^\lambda\rangle \coloneqq \sqrt{\lambda} |0\rangle_R\ox |0\rangle^{\ox k}_{A} + \sqrt{1-\lambda} |1\rangle_R\ox |1\rangle_{A}^{\ox k}.
\label{eq:weighted-repetition-code}
\end{align}
A compact formula for the coherent information of this code in terms of an optimization over the weight $\lambda\in[0,1]$ and arbitrary blocklength $k$ can easily be derived (see \suppref{sec:gadc-codes}).
Note that the optimal single-letter coherent information $\Qone(\cA_{\gamma,N})$ is achieved by \eqref{eq:weighted-repetition-code} with $k=1$ and optimized weight parameter $\lambda\in[0,1]$ \cite{GPPLS09}.
We will show in this section that the neural network state ansatz finds superadditive codes for the GADC that substantially outperform weighted repetition codes.

In the following, we restrict our attention to the interval $N\in[0,1/2]$, as $\cA_{\gamma,N}$ and $\cA_{\gamma,1-N}$ are unitarily equivalent and hence their channel coherent informations (and quantum capacities) coincide \cite{KSW19}.
In the optimization procedure we consider the values $N\in\lbrace 0.1,0.2,0.3,0.4,0.5\rbrace$ and identify intervals of $\gamma$ in which weighted repetition codes are superadditive, that is, they yield a higher coherent information than the optimal single-letter coherent information.
For $k=3,4,5$ copies of $\cA_{\gamma,N}$, we search for neural network codes using a feed-forward architecture as described in Fig.~\ref{fig:networks} with four hidden layers of width $2k$ each.
We choose $\cos$ as the activation function in the first layer, the hyperbolic tangent function $\tanh$ as the activation function in the subsequent layers, and a Cartesian output layer (see \eqref{eq:FF-architecture}).
In contrast to the more common gradient-based optimization techniques in machine learning, we choose to optimize the neural network parameters using stochastic \emph{gradient-free} techniques. 
In particular, we use particle swarm optimization algorithm followed by pattern search.
We motivate our choice to use these algorithms in App.~\ref{sec:numerics}, which also contains high-level explanations of these techniques.

For all values $N\in\lbrace 0.1,0.2,0.3,0.4,0.5\rbrace$ we find neural network codes outperforming the weighted repetition codes \eqref{eq:weighted-repetition-code}, as shown in Fig.~\ref{fig:gadc-codes}. 
For each $N\in\lbrace 0.1,0.2,0.3,0.4,0.5\rbrace$, the codes in Fig.~\ref{fig:gadc-codes} are obtained by first carrying out our optimization technique for a particular value of $\gamma$ close to the threshold of the best weighted repetition code.
We then plot the best neural network code found in this manner for the entire interval $\gamma$ where superadditivity occurs.
As a benchmark, we evaluate weighted repetition codes for up to $k=16$ channel copies using the formula derived in \suppref{sec:gadc-codes}; the codes $\phi_k$ for $1\leq k\leq 5$ perform best and are shown in Fig.~\ref{fig:gadc-codes} for comparison.

\begin{figure}
	\centering
	\includegraphics[width=\textwidth]{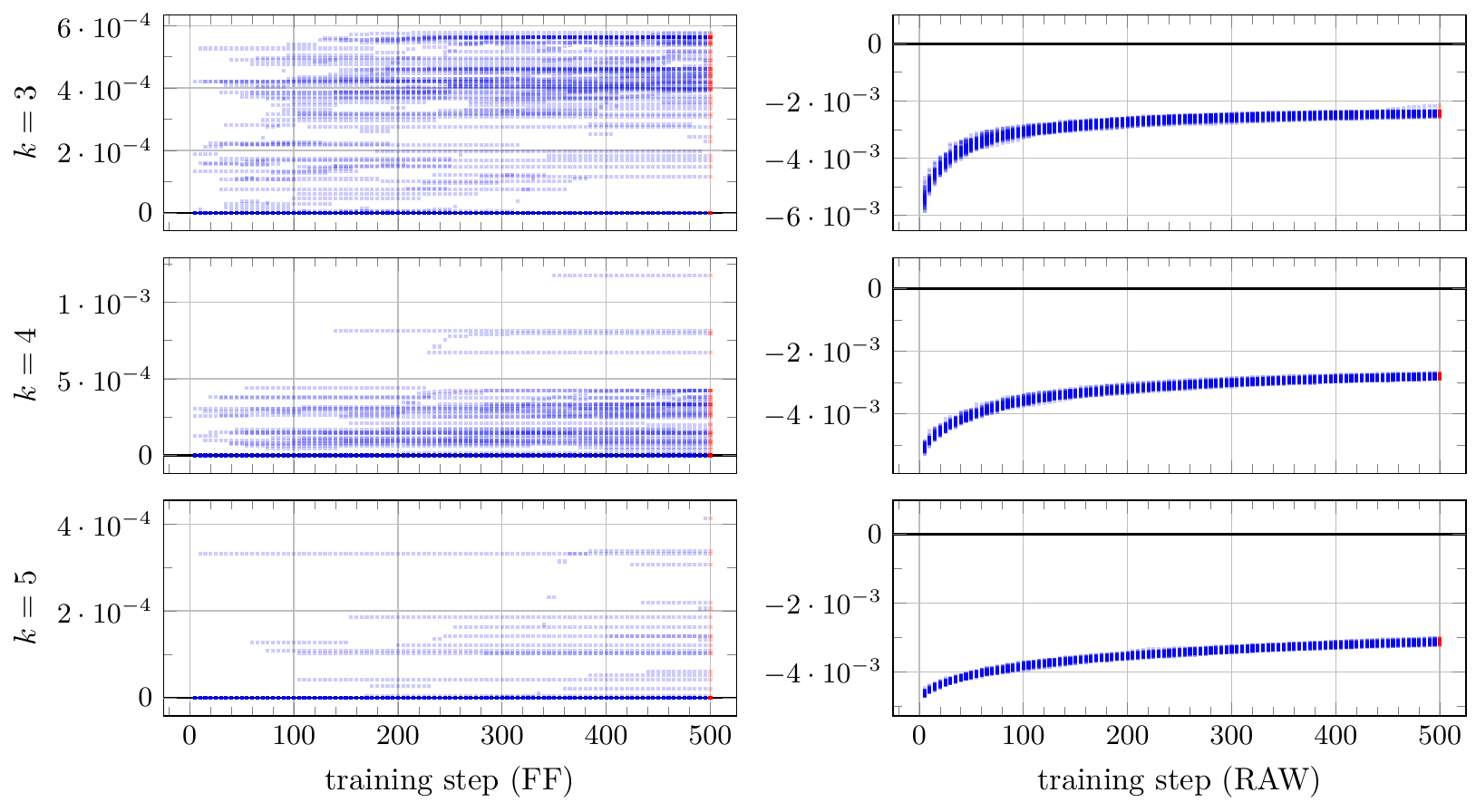}
	\caption{Training convergence of a particle swarm optimization algorithm maximizing the CI of $k=3,4,5$ copies of the generalized amplitude damping channel $\cA_{\gamma,N}$ with parameters $(\gamma,N) = (0.44035,0.1)$. The left column plots a feed-forward (FF) net representation with four hidden layers of width $2k$ each (see Sec.~\ref{sec:gadc}), having 182/306/462 real parameters for $k=3,4,5$, respectively. The right column plots a direct parametrization (RAW) of the $2^{(2k)}$ complex amplitudes, resulting in 64/256/1024 real parameters for $k=3,4,5$, respectively.}
	\label{fig:gadc-pso-convergence}
\end{figure}

We focus here on the neural network codes found for the values $(\gamma,N) = (0.44035,0.1)$ and $k=3,4,5$ copies of $\cA_{\gamma,N}$, and note that the neural network codes for the other values of $(\gamma,N)$ are collected in App.~\ref{sec:gadc-codes}.
In Tab.~\ref{tab:gadc-codes-N01} we list the best codes (as plotted in Fig.~\ref{fig:gadc-codes}) for each blocklength together with their coherent information.
In Fig.~\ref{fig:gadc-pso-convergence} we plot the convergence of the particle swarm optimization algorithm for $(\gamma,N) = (0.44035,0.1)$ and $k=3,4,5$ (FF), and compare its performance to a direct parametrization (RAW) of the $2^{2k}$ complex amplitudes in the quantum code $|\psi_n\rangle$, again optimized using PSO.
Evidently, using comparable optimization parameters the raw ansatz is not able to find even trivial product codes with coherent information equal to zero.
Note also that for $(\gamma,N) = (0.44035,0.1)$ the weighted repetition codes in \eqref{eq:weighted-repetition-code} do not yield positive coherent information up to at least $k=16$.
Hence, the neural network codes increase the \emph{threshold} of the GADC substantially, as seen in Fig.~\ref{fig:gadc-codes}.
The threshold of a parametrized family of quantum channels is defined as the boundary of the region in which the channel has positive quantum capacity.

\begin{table}
	\centering
	\begin{tabular}{llll}
		\toprule
	$|\nu_k\rangle$ & $s^n$ ($A^k|R$) & $\psi(s^n)$ & $\frac{1}{k}\Qone(\nu_k,\cA_{\gamma,N}^{\ox k})$\\
	\midrule 
	$k=3$ 
	& \phantom{$00$}$000|000$\phantom{$00$} & $-0.3934 + 0.2231i$& $5.7598\cdot 10^{-4}$\\
	& \phantom{$00$}$000|110$\phantom{$00$} & $-0.3136 + 0.2501i$&\\
	& \phantom{$00$}$001|111$\phantom{$00$} & $-0.3956 + 0.2345i$&\\
	& \phantom{$00$}$010|111$\phantom{$00$} & $-0.3956 + 0.2346i$&\\
	& \phantom{$00$}$100|111$\phantom{$00$} & $-0.3955 + 0.2348i$&\\
	\midrule
	$k=4$ 
	& \phantom{0}$0101|1110$\phantom{0} & $+0.3482 - 0.2537i$& $1.2683 \cdot 10^{-3}$\\
	& \phantom{0}$1010|1110$\phantom{0} & $+0.3354 - 0.2723i$&\\
	& \phantom{0}$1111|0001$\phantom{0} & $+0.3986 - 0.3920i$&\\
	& \phantom{0}$1111|1000$\phantom{0} & $+0.3980 - 0.3959i$&\\ 
	\midrule
	$k=5$ 
	& $01010|00011$ &  $+0.3010 + 0.1666i$& $9.1537 \cdot 10^{-4}$\\
	& $10101|00011$ &  $+0.4389 + 0.1233i$&\\
	& $11111|10110$ &  $+0.5660 + 0.0816i$&\\
	& $11111|11101$ &  $+0.5844 + 0.0725i$&\\
	\bottomrule
	\end{tabular}
\caption{Table of the best neural network codes for the GADC $\cA_{\gamma,N}$ with $(\gamma,N) = (0.44035,0.1)$ and $k=3,4,5$ channel copies. Only the non-zero amplitudes $\psi(s^n)$ indexed by the basis string $s^n$ (with $n =2k$) are shown (see \eqref{eq:qudit-state-main}). The architecture used for the neural network codes is a feed-forward net with four hidden layers of width $2k$ each, activation functions $\cos$ and $3\times \tanh$, and a Cartesian output layer (see Sec.~\ref{sec:gadc}).}
\label{tab:gadc-codes-N01}
\end{table}

\subsection{Dephrasure Channel}\label{sec:dephrasure-main}
The neural network ansatz is also able to find new quantum codes for the \emph{dephrasure channel} that was introduced recently in \cite{LLS18}.
It is defined in terms of probabilities $p,q\in[0,1]$ as
\begin{equation}
\cN_{p,q}(\rho) = (1-q)((1-p)\rho + p Z\rho Z) + q \tr(\rho) |e\rangle\!\langle e|,\label{eq:dephrasure}
\end{equation}
where $Z=|0\rangle\!\langle 0|-|1\rangle\!\langle 1|$ is the Pauli $Z$-operator, and $|e\rangle$ is an erasure flag that is orthogonal to the input space.
The name `dephrasure' is derived from the fact that $\cN_{p,q}$ first dephases an input state in the $Z$-basis with probability $p$, and then erases it with probability $q$.
Despite the fact that both dephasing and erasure noise are well-understood in terms of quantum information transmission, the dephrasure channel---a concatenation of the two---exhibits superadditivity of coherent information for as little as two uses of the channel \cite{LLS18}.
As a result, the quantum capacity of the dephrasure channel is unknown for a large region in the parameter space.

As for the GADC in the previous section, superadditivity of coherent information for the dephrasure channel is again achieved by weighted repetition codes $\phi^\lambda_k$ as defined in \eqref{eq:weighted-repetition-code}.
A compact formula for the coherent information $\frac{1}{k} \Qone(\phi^\lambda_k,\cN_{p,q}^{\ox k})$ of these codes was derived in \cite{LLS18}, and we state it in \suppref{sec:dephrasure-codes}.
Similar to the GADC in Sec.~\ref{sec:gadc}, we note that the optimal single-letter coherent information $\Qone(\cN_{p,q})$ for the dephrasure channel is achieved by $\phi_1^\lambda$ for some $\lambda\in[0,1]$ \cite{LLS18}.
We show in this section that the neural network state ansatz finds new quantum codes demonstrating even larger superadditivity of coherent information for the dephrasure channel.

In the following, we focus our attention to the values $q\in\lbrace 0.1,0.2,0.3,0.4\rbrace$ of the erasure probability; for each $q$, we then investigate values of the dephasing probability $p$ for which weighted repetition codes achieve superadditivity.
Since the dephrasure channel maps a qubit to a qutrit, optimizing its coherent information is computationally more costly than for the GADC, which forces us to restrict our attention to $k=2,3,4$ copies of $\cN_{p,q}$ (we refer to Sec.~\ref{sec:discussion} for a discussion of these numerical limitations).
We again use a feed-forward network as described in Fig.\ \ref{fig:networks} with four hidden layers of width $2k$ each and $\cos$ as the activation function in the first layer.
However, in contrast to Sec.~\ref{sec:gadc} we use $\relu$ as the activation function in the remaining layers, and an exponential output layer corresponding to a polar parametrization instead of a Cartesian one.
We found these choices to perform significantly better for the dephrasure channel.
As in Sec.~\ref{sec:gadc}, the neural network parameters were optimized using the particle swarm optimization algorithm followed by pattern search (see App.~\ref{sec:numerics}).

\begin{figure}
	\centering
	\includegraphics[width=\textwidth]{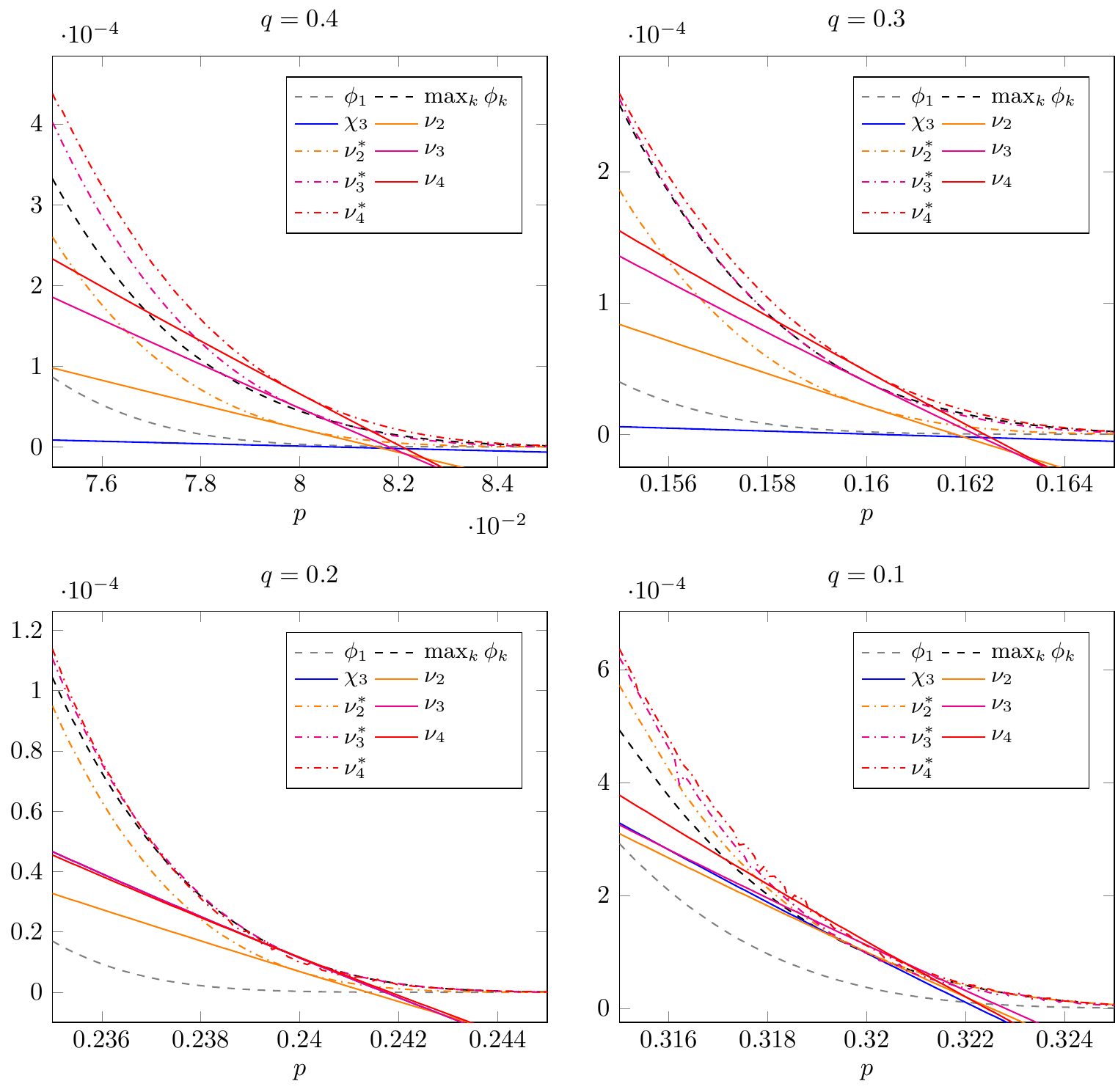}
	\caption{Overview of quantum codes for the dephrasure channel $\cN_{p,q}$ comparing the neural network codes $\nu_k$ (solid orange, magenta, and red lines) for $k=2,3,4$ to the optimal single-letter code $\phi_1$ (grey dashed line) and the maximum over all weighted repetition codes $\phi_k$ (black dashed line) for $2\leq k\leq 10$ defined in \eqref{eq:weighted-repetition-code}.	
	We also plot the neural network codes $\nu_k^*$ with optimized parameters over the shown interval (dash-dotted lines), and the best code $\chi_3$  on three channel qubits found with a direct parametrization of the quantum state amplitudes (blue line).
	For each $q\in\lbrace 0.1,0.2,0.3,0.4\rbrace$, we plot the interval of $p$ in which superadditivity occurs. 
	The neural network codes $\nu_k$ are listed in Tabs.~\ref{tab:dephr-codes-q04}, \ref{tab:dephr-codes-q03}, \ref{tab:dephr-codes-q02}, \ref{tab:dephr-codes-q01} for $q=0.4,0.3,0.2,0.1$, respectively.
	}
	\label{fig:dephr-codes}
\end{figure}

For all values $q\in\lbrace 0.1,0.2,0.3,0.4\rbrace$ we find neural network codes outperforming the weighted repetition codes \eqref{eq:weighted-repetition-code}, as shown in Fig.~\ref{fig:dephr-codes}. 
For each $q\in\lbrace 0.1,0.2,0.3,0.4\rbrace$ and $k=2,3,4$, the codes in Fig.~\ref{fig:dephr-codes} are obtained by first carrying out our optimization technique for a particular value of $p$ close to the threshold of the best weighted repetition code.
We then plot the best neural network code (labeled $\nu_k$ for $k=2,3,4$) found in this manner for an interval of $p$ where superadditivity occurs.
We also individually optimized the coefficients of the basis strings $s^n$ with non-zero weight across the shown interval of $p$, yielding even better codes $\nu^*_k$.
Curiously, such an additional optimization over coefficients gave no improvement for the neural network codes found for the GADC in Sec.~\ref{sec:gadc}.
In contrast, for fixed $q$ there is an evident interplay between the dephasing probability $p$ in the dephrasure channel $\cN_{p,q}$ and the coefficients of the neural network codes $\nu_k$, as evident from Fig.~\ref{fig:dephr-codes}.
As a benchmark, we evaluated weighted repetition codes $\phi_k$ for up to $k=10$ channel copies using the formula in App.~\ref{sec:dephrasure-codes}; the maximum $\max_k\phi_k$ over these codes for $1\leq k\leq 10$ is shown in Fig.~\ref{fig:dephr-codes} for comparison, along with the optimal single-letter code $\phi_1$.

\begin{figure}
	\centering
	\includegraphics[width=\textwidth]{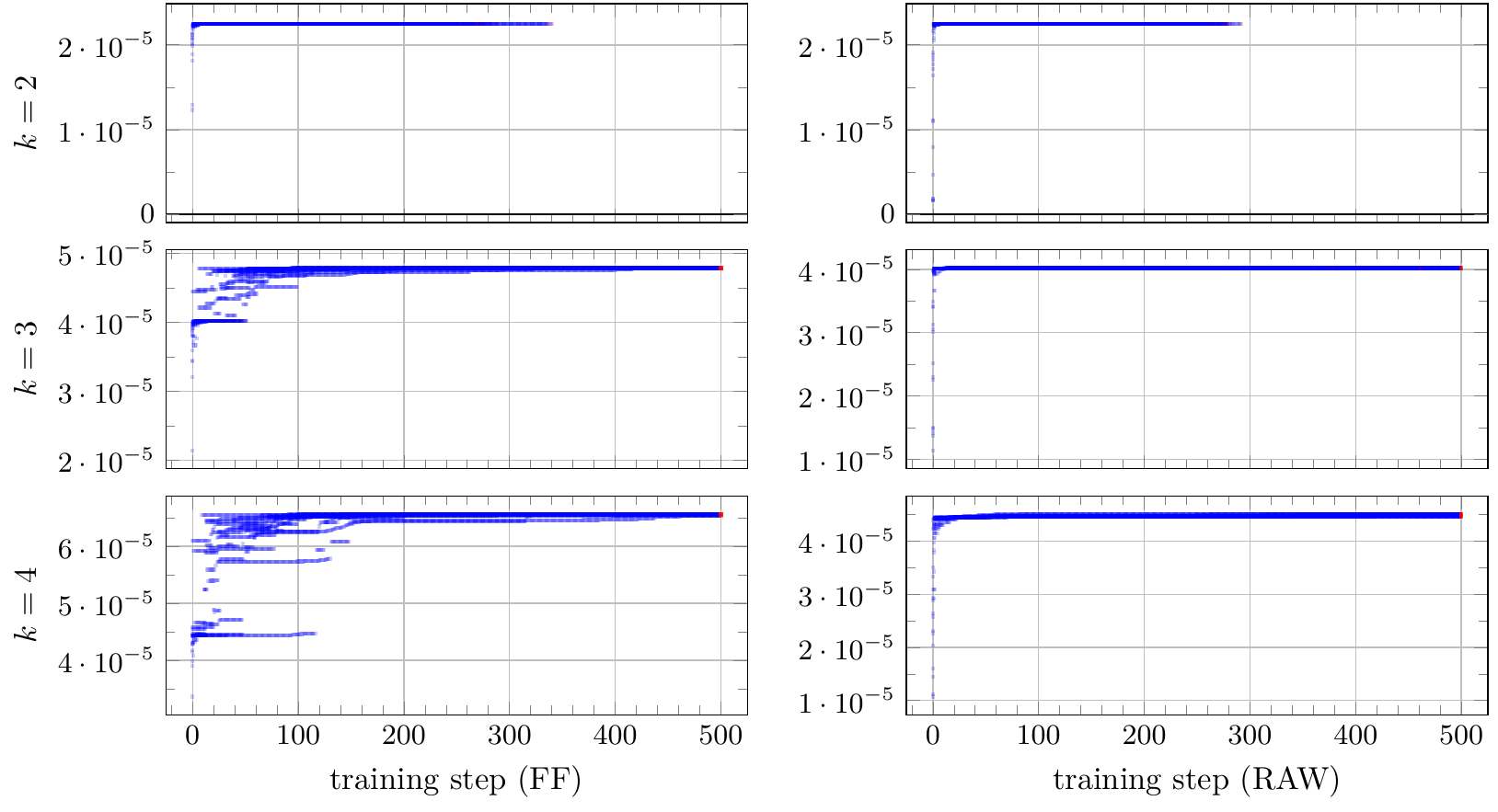}
	\caption{Training convergence of a particle swarm optimization algorithm maximizing the CI of $k=2,3,4$ copies of the dephrasure channel $\cN_{p,q}$ with parameters $(p,q) = (0.08,0.4)$. 
		The left column plots a feed-forward (FF) net representation with four hidden layers of width $2k$ each (see Sec.~\ref{sec:dephrasure-main}), having 90/182/306 real parameters for $k=2,3,4$, respectively. 
		The right column plots a direct parametrization (RAW) of the $2^{(2k)}$ complex amplitudes, resulting in 32/128/512 real parameters for $k=2,3,4$, respectively.
	While the two parametrization find equivalent codes for $k=2$, the feed-forward net representation finds strictly better codes for $k=3,4$ than the raw parametrization.}
	\label{fig:dephr-pso-convergence}
\end{figure}

We focus in the following on the neural network codes found for the values $(p,q)=(0.08,0.4)$ and $k=2,3,4$; the other neural network codes are listed in App.~\ref{sec:dephrasure-codes}.
In Tab.~\ref{tab:dephr-codes-q04} we list the best codes (as plotted in Fig.~\ref{fig:dephr-codes}) for each blocklength together with their coherent information.
In Fig.~\ref{fig:dephr-pso-convergence} we plot the convergence of the particle swarm optimization algorithm for $(p,q) = (0.08,0.4)$ and $k=2,3,4$ (FF), and compare its performance to a direct parametrization (RAW) of the $2^{2k}$ complex amplitudes in the quantum code $|\psi_n\rangle$, again optimized using PSO.
Similar to the GADC in Sec.~\ref{sec:gadc}, the raw ansatz is not able to find codes with coherent information rates as high as the neural network codes.
However, in contrast to the GADC the raw ansatz is indeed able to find superadditive quantum codes.
For $k=2$, these codes found using the raw ansatz are optimal (as already observed in \cite{LLS18}), while for $k=3,4$ they are clearly outperformed by our neural network codes.
Another observation of \cite{LLS18} is that the dephasing part of $\cN_{p,q}$ suggests a Schmidt ansatz for quantum codes, a neural network state version of which is discussed in \eqref{eq:qubit-state-schmidt} in Sec.~\ref{sec:depolarizing}.
However, in the high-noise regime investigated above, this Schmidt ansatz did not yield codes performing as well as the codes $\nu_k$ resp.~$\nu_k^*$.

\begin{table}
	\centering
	\begin{tabular}{llll}
		\toprule
		$|\nu_k\rangle$ & $s^n$ ($A^k|R$) & $\psi(s^n)$ & $\frac{1}{k}\Qone(\nu_k,\cN_{p,q}^{\ox k})$\\
		\midrule 
		$k=2$ 
		& \phantom{$00$}$00|00$\phantom{$00$} & $-0.2504-0.4352i$ & $2.2502\cdot 10^{-5}$\\
		& \phantom{$00$}$00|01$\phantom{$00$} & $-0.6941+0.5142i$&\\
		& \phantom{$00$}$11|01$\phantom{$00$} & $+0.0374+0.0171i$&\\
		& \phantom{$00$}$11|11$\phantom{$00$} & $-0.0001 - 0.0001i$&\\
		\midrule
		$k=3$ 
		& \phantom{0}$000|011$\phantom{0} & $-0.0304+0.0465i$& $4.7881 \cdot 10^{-5}$\\
		& \phantom{0}$001|011$\phantom{0} & $-0.0465+0.0408i$&\\
		& \phantom{0}$111|000$\phantom{0} & $-0.6954+0.7138i$&\\
		\midrule
		$k=4$ 
		& $0101|1000$ &  $+0.0022-0.0031i$& $6.5699 \cdot 10^{-5}$\\
		& $0111|1000$ &  $-0.9169-0.3686i$&\\
		& $0111|1110$ &  $+0.0054+0.0102i$&\\
		& $1000|0111$ &  $-0.0932+0.0049i$&\\
		& $1001|0111$ &  $+0.0001+0.0000i$&\\
		& $1010|0111$ &  $-0.0341-0.1156i$&\\
		\bottomrule
	\end{tabular}
	\caption{Table of the best neural network codes for the dephrasure channel $\cN_{p,q}$ with $(p,q) = (0.08,0.4)$ and $k=2,3,4$ channel copies. Only the non-zero amplitudes $\psi(s^n)$ indexed by the basis string $s^n$ (with $n =2k$) are shown (see \eqref{eq:qudit-state-main}). The architecture used for the neural network codes is a feed-forward net with four hidden layers of width $2k$ each, activation functions $\cos$ and $3\times \relu$, and a Polar output layer (see Sec.~\ref{sec:dephrasure-main}).}
	\label{tab:dephr-codes-q04}
\end{table}

\section{Representing the Best Known Codes for the Depolarizing Channel}\label{sec:depolarizing}
The \emph{depolarizing channel} is used as a model to describe qubit decoherence in a noisy environment.
For a qubit in a state described by the density operator $\rho$, and for a real parameter $p\in[0,4/3]$, the action of the channel is given by 
\begin{equation}\label{eq:depolarizing}
\cD_p(\rho) \coloneqq (1-p)\rho+p\tr(\rho)(|0\rangle\!\langle0|+|1\rangle\!\langle1|)/2,
\end{equation}
i.e.\ the original state $\rho$ is replaced by the maximally mixed state with `probability' $p$ (for $p\le1$); in other words, if on the Bloch sphere $\rho$ has spin polarization vector $\vec x$, the channel $\cD_p$ shrinks $\vec x$ by a factor $1-p$.

For the depolarizing channel, the single-letter channel coherent information $\Qone(\cD_p)$ is maximized by a Bell state $\frac{1}{\sqrt{2}}(|0\rangle_R |0\rangle_A + |1\rangle_R|1\rangle_A)$, and evaluates to \cite{Wil13}
\begin{align}
Q^{(1)}(\cD_p) = 1 + \left(1-\frac{3p}{4}\right)\log\left(1-\frac{3p}{4}\right) + \frac{3p}{4}\log\frac{p}{4}.\label{eq:single-letter-cohinfo}
\end{align}
$Q^{(1)}(\cD_p)$ remains positive up to the threshold at $p=0.25238$ (the \emph{threshold} is defined as the highest $p$ for which $\Qone(\cD_p) > 0$).
The next highest thresholds are achieved for $k=3$ and $5$ channel copies and a $k$-repetition code 
\begin{align}
|\phi_k\rangle=\frac{1}{\sqrt{2}}(|0\rangle_R|0\rangle_A^{\ox k} + |1\rangle_R|1\rangle_A^{\ox k}),\label{eq:repetition-code}
\end{align} 
for which the channel coherent information $\Qone(\phi_k,\cD_p^{\ox k})$ reaches zero at $p=0.25350$ and $p=0.25380$, respectively.
Both in terms of the rate and the threshold, these repetition codes are the best known codes up to 9 channel copies, which is discussed in more detail in \suppref[App]{sec:product-rep-codes}.

We show in the following that a variational neural network ansatz achieves these codes for the depolarizing channel.
We also contrast the various architectures (RBM, feed-forward, and their Schmidt variants) on an empirical level.
To compute the amplitude function $\psi(i^n)$ in the tensor basis expansion
\begin{align}
|\psi_{n}\rangle = \frac{1}{C} \sum_{i^n\in \lbrace 0,1\rbrace^n} \psi(i^n) |i^n\rangle \in (\mathbb{C}^2)^{\ox n},\label{eq:qudit-state-main}
\end{align}
we use both an RBM architecture as well as an FF architecture with a $\cos$ activation function in the first hidden layer, and $\relu$ in two subsequent hidden layers.
This setup, which has been shown to perform well in the context of representing quantum states of local Hamiltonians \cite{CL18}, clearly outperformed a $\relu$-only architecture in our numerical investigations of the GADC and the dephrasure channel in Sec.~\ref{sec:new-codes}.

Furthermore, we propose a Schmidt-ansatz similar to \eqref{eq:qudit-state-main} given for $2l$ qubits by
\begin{align}
|\psi_{2l}\rangle = \frac{1}{C} \sum_{i^l\in \lbrace 0,1\rbrace^l} \psi(i^l) |i^l\rangle_R|i^l\rangle_A.\label{eq:qubit-state-schmidt}
\end{align}
This approach greatly reduces the number of degrees of freedom required to parametrize $|\psi_{2l}\rangle$, but enforces the environment $R$ to have the same dimension as the system $A$.
Note that this may introduce redundancy, as e.g.\ a repetition code ordinarily only requires a single purifying qubit.
The ansatz in \eqref{eq:qubit-state-schmidt} furthermore introduces a choice of basis for the channel input qubits, rendering it less general than the ansatz in \eqref{eq:qudit-state-main}.

Using an explicit construction, we show that both FF and RBM architectures can efficiently represent products of repetition codes (which are discussed in App.~\ref{sec:analytical-proofs}):
given $k$ repetition codes on $n_1,\ldots,n_k$ qubits, respectively, an RBM with $\prod_in_i$ visible units and $k$ hidden nodes can represent the corresponding state amplitudes, and a FF net with first $\cos$ and second $\relu$ hidden layer width $k$, and a single final $\relu$ node suffices.

\begin{figure}
	\centering
	\includegraphics[width=1\textwidth]{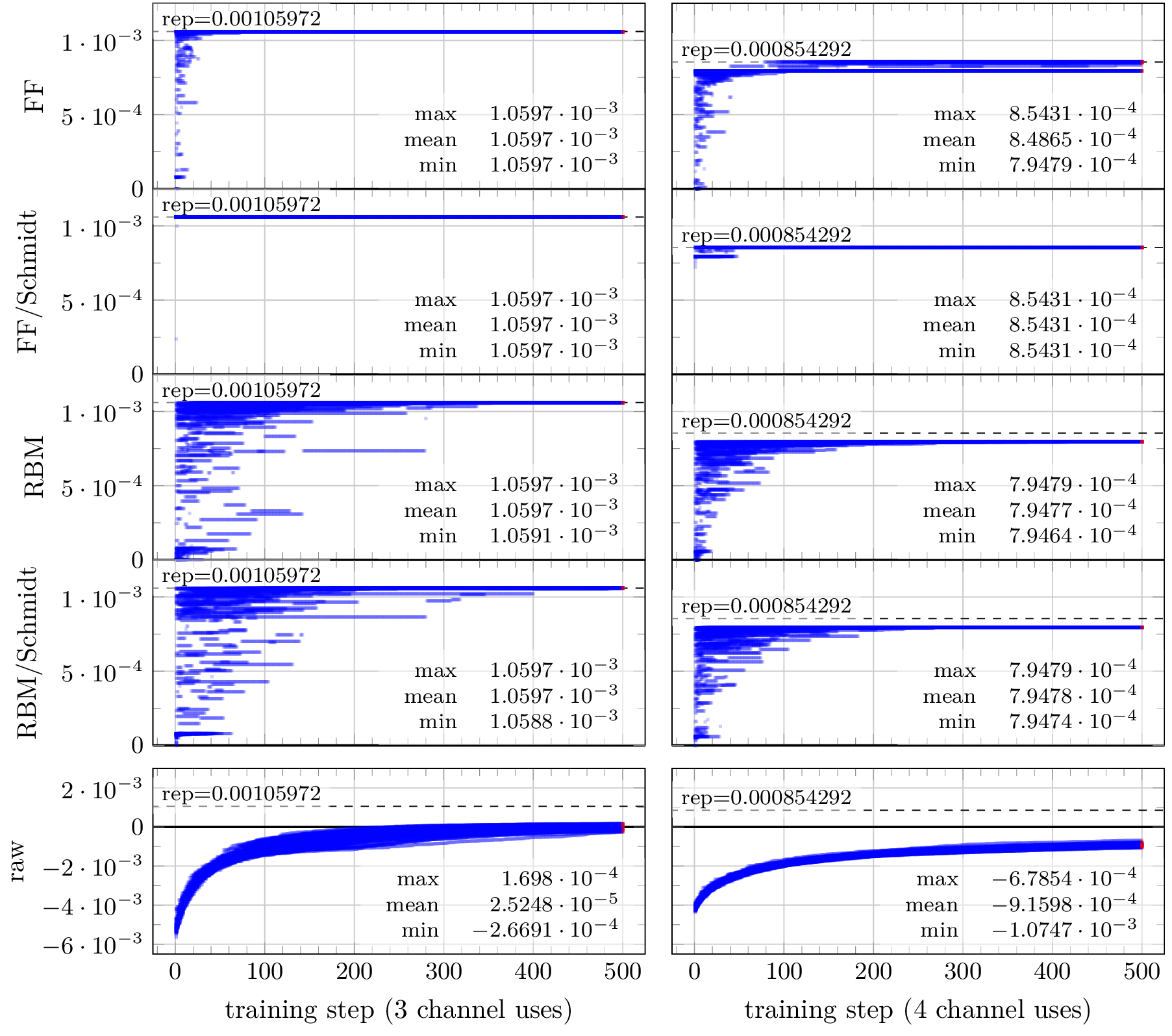}
	\caption{Training convergence of a particle swarm algorithm maximizing the CI of three resp.\ four copies of the depolarizing channel $\cD_p$, with noise parameter $p=0.2523$.
		Plotted are the best candidates of $80$ threads \`a 100 particles for every training step from 0 to 500.
		The final candidate distribution, and the outcome of other optimization algorithms can be seen in \suppref{sec:extra-numerics}.
		For three channel copies, a three-repetition code maximizes the coherent information, whereas for four channel copies a product code of a three-repetition and single-repetition code is optimal.
		Plotted are FF (feed-forward net, 140 resp.\ 234 real parameters; see Sec.~\ref{sec:depolarizing} for the FF architecture), FF/Schmidt (Schmidt representation obtained from a feed-forward net, 40 resp.\ 65 real parameters), RBM (restricted Boltzmann machine with hidden layer width 9, 138 resp.\ 232 real parameters), RBM/Schmidt (Schmidt representation obtained from an RBM with hidden layer width 9, 39 resp.\ 64 real parameters), and raw (parametrizing the full state vector, 128 resp.\ 512 real parameters); note that the FF and RBM representations are in fact overspecified for three channel uses.}
	\label{fig:convergence-dep}
\end{figure}

Empirically, we contrast FF, RBM and their corresponding Schmidt variants as a variational ansatz $\psi_n$ (with $n=2k$) to maximize $\Qone(\psi_n,\cD_p^{\otimes k})$; the FF architecture consists of three hidden layers of width $n=2k$ with $\cos$-$\relu$-$\relu$ for the activation functions and a Cartesian output layer.
In comparison with a full state vector on $n$ qubits with $2\times2^n$ real parameters, we can see a significant improvement in convergence speed (see Fig.\ \ref{fig:convergence-dep}), both in the case that the best-known code is a single repetition code for three channel uses, or a three times one product repetition code (see \suppref{sec:product-rep-codes} for an explanation of this terminology).
For both FF and RBM architectures, the Schmidt ansatz \eqref{eq:qubit-state-schmidt} surpasses the standard parametrization \eqref{eq:qudit-state-main}, which is likely due to the significantly-reduced parameter count.
FF networks further outperform RBM architectures with comparable parameter counts on three and four channel uses of a depolarizing channel, which we verified with various global derivative-free optimization techniques (see \suppref{sec:numerics} for an overview) to reduce the likelihood of a systematic bias in our numerical findings.
The numerical data for these findings is collected in \suppref{sec:extra-numerics}.
We also note that a deep Boltzmann machine ansatz as described in Sec.~\ref{sec:1-dbm-states} offered no advantage over an RBM ansatz, neither in terms of representability nor convergence speed.

\section{Representing Absolutely Maximally Entangled States}\label{sec:ame-main}
Absolutely maximally entangled (AME) states are $n$-partite states having maximal correlation across any bipartition of the $n$ parties into equal halves.
These states are certain examples of quantum error-correcting codes, whose intricate multipartite entanglement structure mediates correlations between different subsets of the constituent systems. 

AME states can be used as a resource for multi-user information-theoretic tasks such as open-destination teleporation, secret sharing or entanglement swapping that require maximal entanglement across different choices of bipartitions \cite{HC13,HCLRL12}.
In a holographic context, where AME states are referred to as \emph{perfect tensors}, they provide examples of holographic error-correcting codes \cite{LS15,PYHP15,LHGZ17}.
More generally, an arbitrary AME state on $n$ qudits of local dimension $d$ can be interpreted as a $((n,1,\lfloor\frac{n}{2}\rfloor+1))_d$ quantum error-correcting code, i.e., a code of distance $\lfloor\frac{n}{2}\rfloor+1$ encoding a 1-dimensional subspace in $n$ physical qudits \cite{Sco04}.

To define absolutely maximally entangled (AME) states in a precise way, we consider a pure state $|\psi_{n,d}\rangle\in(\mathbb{C}^d)^{\ox n}$ on $n$ qudits of local dimension $d$.
For a subset $\cS\subset[n]\coloneqq \{ 1,\dots,n\}$ of the $n$ qudits we denote by $\rho_\cS = \tr_{\cS^c}\psi_{n,d}$ the marginal of $\psi_{n,d}$ on $\cS$. 
Then $\psi_{n,d}$ is AME if $\rho_\cS = \frac{1}{|\cS|} I_\cS$ for every $\cS\subset[n]$ with $|\cS|=\lfloor \frac{n}{2}\rfloor$.
We use the notation $\ame(n,d)$ for an AME state on $n$ qudits of local dimension $d$.

Since an AME state is maximally entangled across all possible bipartitions into equal halves, monogamy of entanglement \cite{CKW00} puts an obstruction on their existence. 
Furthermore, the fact that AME states are particular quantum error-correcting codes yields additional constraints via weight enumerator theory \cite{SL97,Rai98}. 
Consequently, AME states do not exist for all $(n,d)$ \suppref[Sec]{sec:ame}.
For example, it is known that there is no $\ame(4,2)$ state \cite{HS00}.
On the other hand, an example of an $\ame(4,3)$ state is $|\Omega_{4,3}\rangle = \frac{1}{3} \sum_{i,j=0,1,2} |i\rangle |j\rangle |i+j (3)\rangle |i+2j(3)\rangle,$ where $k(d)\equiv k\mod d$.

The property of $\psi_{n,d}$ being absolutely maximally entangled is related to the \emph{linear entropy} $S_L(\rho_\cS) = \frac{d^m}{d^m-1}(1-\tr(\rho_\cS^2))$ of the marginals $\rho_\cS$ for $\cS\subset[n]$ with $|\cS| = \lfloor\frac{n}{2}\rfloor$.
Defining for $m=1,\dots,\lfloor\frac{n}{2}\rfloor$ the average linear entropy
\begin{align}
Q_m(\psi_{n,d})\coloneqq \binom{n}{m}^{-1}\sum_{\cS\subset[n]\colon |\cS|=m} S_L(\rho_S),\label{eq:Qm}
\end{align}
a pure state $\psi_{n,d}$ is AME if and only if $Q_{\lfloor \frac{n}{2}\rfloor}(\psi_{n,d}) = 1$ \cite{Sco04}.
Hence, to search for $\ame(n,d)$-states $\psi_{n,d}$, we can use \eqref{eq:Qm} with $m=\lfloor \frac{n}{2}\rfloor$ as the objective function and optimize the parameters in an ansatz for $\psi_{n,d}$ such that $Q_{\lfloor \frac{n}{2}\rfloor}(\psi_{n,d}) \approx 1$.
As before, we use a neural network state ansatz for $\psi_{n,d}$ based on the following decomposition with respect to a given basis $\lbrace |i\rangle\rbrace_{i=0}^{d-1}$:
\begin{align}
|\psi_{n,d}\rangle = \frac{1}{C} \sum_{i^n\in [d]_0^n} \psi(i^n) |i^n\rangle,
\end{align}
where as before $C$ is a normalization constant, and we use the notation $[d]_0 = \lbrace 0,\dots,d-1\rbrace$.
The amplitude function $\psi(i^n)$ is again computed by a neural network; since this is now a function from the set of all $d$-ary strings of length $n$ into $\mathbb{C}$, there are multiple options how to encode $i^n$ as the input to a neural network.
We discuss these options in detail in \suppref{sec:encodings}.

\begin{figure}[t]
	\includegraphics[width=\textwidth]{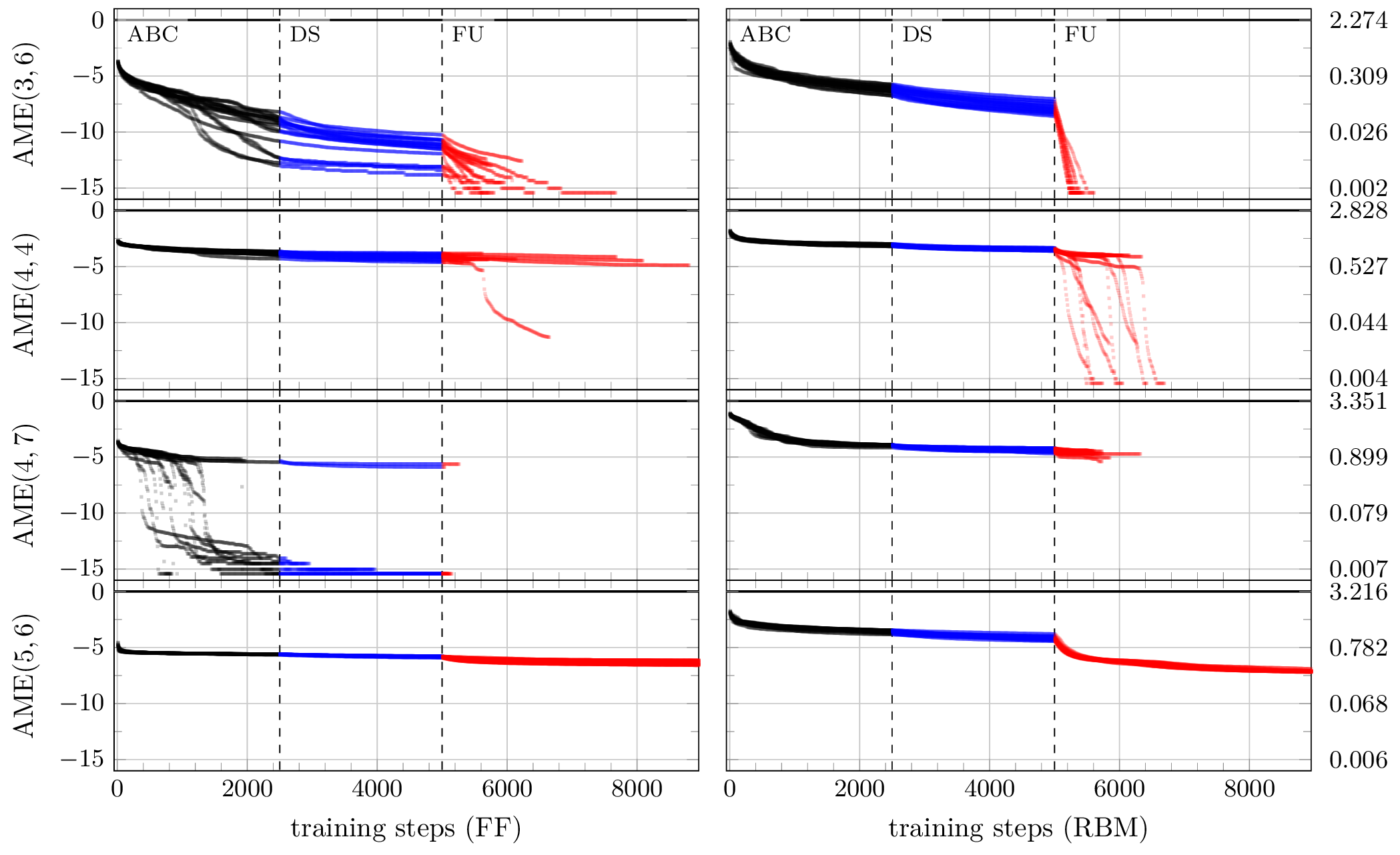}
	\caption{Training convergence for representing $\ame(n,d)$-states for the cases $(n,d) \in \lbrace (3,6), (4,4), (4,7), (5,6)\rbrace$ using consecutive steps of artifical bee colonization (black), direct search (blue), and gradient search (red).
		On the left y-axis we plot the quantity $\ln(1-Q_m)$, and on the right y-axis we plot the bound in \eqref{eq:trace-parameter-bound} on the average trace distance parameter $D_m$ defined in \eqref{eq:AME-dist}.
		The encodings used to represent the $d$-ary input strings, as well as the network architectures for the FF and RBM nets are listed in Tab.~\ref{tab:ame-architecture}.}
	\label{fig:ame-states}
\end{figure}

We demonstrate in Fig.~\ref{fig:ame-states} that parametrizing $\psi_{n,d}$ with a neural network state ansatz yields $\ame(n,d)$-states for the pairs $(n,d)=(3,6)$, $(4,4)$, $(4,7)$, and $(5,6)$.
For the numerical optimization, we use the artificial bee colonization algorithm, followed by pattern search and a final round of gradient search (see \suppref{sec:numerics}).
These choices of parameters are only exemplary, and the neural network state ansatz is capable of representing $\ame(n,d)$-states also for other pairs $(n,d)$ such as $(3,3)$, $(4,3)$, and $(4,5)$.
In the last three cases, the convergence is remarkably fast and only takes a few iterations in optimization algorithms such as ABC or PSO to reach a value of $Q_{\lfloor \frac{n}{2}\rfloor}$ sufficiently close to 1.

To assess our numerical results, we introduce an `average trace distance' parameter 
\begin{align}
D_m(\psi_{n,d}) \coloneqq \binom{n}{m}^{-1} \sum_{\cS\subset[n]\colon |\cS|=m} \left\|\rho_S - \pi_\cS \right\|_1, \label{eq:AME-dist}
\end{align}
where $\pi_\cS\coloneqq\frac{1}{|\cS|} I_\cS$ denotes the completely mixed state, and $\|X\|_1=\tr\sqrt{X^\dagger X}$ is the trace norm of an operator $X$.
The parameter $D_m(\psi_{n,d})$ measures the average trace distance of the marginals of a state $\psi_{n,d}$ on $m$ subsystems to the completely mixed state.
Clearly, $D_{\lfloor\frac{n}{2}\rfloor}(\psi_{n,d}) = 0$ if and only if $\psi_{n,d}$ is AME.
We prove in \suppref[Sec]{sec:ame} that 
\begin{align}
D_m(\psi_{n,d}) \leq \sqrt{2\log[d^m-(d^m-1)Q_m(\psi_{n,d})]}.\label{eq:trace-parameter-bound}
\end{align}
This bound allows us to relate a value of $Q_m$ to how close (on average) in trace distance a state is to being AME (see Fig.~\ref{fig:ame-states}).

\begin{table}
	\centering
	\renewcommand{\arraystretch}{1.2}
	\begin{tabular}{lllll}
	\toprule
	$n$ & $d$ & architecture & encoding & hidden layer\\
	\midrule
	$3$ & $6$ & FF & binary & $(12,12,12)$ \\
	& & RBM & binary & $12$\\
	$4$ & $4$ & FF & binary & $(8,8,8)$ \\
	& & RBM & binary & $8$\\
	$4$ & $7$ & FF & scaled & $(4,4,4)$\\
	& & RBM & scaled & $4$\\
	$5$ & $6$ & FF & binary & $(15,15,15,15)$ \\
	& & RBM & binary & $20$ \\
	\bottomrule
\end{tabular}
	\caption{Encodings and network architectures used to represent $\ame(n,d)$-states as plotted in Fig.~\ref{fig:ame-states}. For the FF nets, the hidden layers are denoted by $(M_i)_i$, where $M_i$ is the width of the $i$-th hidden layer. We always use $\cos$ in the first hidden layer, $\relu$ in the following hidden layers, and a Cartesian output layer. For the RBM net, we list the width of the single hidden layer. The encodings are defined in \suppref[Tab]{tab:encodings}.}
	\label{tab:ame-architecture}
\end{table}

\section{Discussion}\label{sec:discussion} 
In this work, we have shown that quantum codes for noisy quantum communication and certain quantum error-correcting codes can be modeled efficiently with various neural network representations.
In particular, we investigated quantum codes that yield high coherent information for the generalized amplitude damping channel (GADC), the dephrasure channel, and the depolarizing channel.
For the GADC and the dephrasure channel, the neural network ansatz finds codes that outperform the best known codes found with traditional numerical methods.
For $k\leq 6$ of the depolarizing channel, we analyzed the representative power of neural network states with regards to the best known codes, repetition codes, and benchmarked how well they can be trained using a variety of global optimization algorithms. 
Finally, we demonstrated how neural network states can represent absolutely maximally entangled states on $n$ qudits of local dimension $d$ for an array of pairs $(n,d)$.

An interesting question is, of course, whether a neural network state ansatz can be used to find better quantum codes for the depolarizing channel in the high noise regime: either in terms of a higher rate than, say, the 5-repetition code right below the noise threshold, or in terms of increasing the noise threshold itself.
Our results indicate that in order to find such codes outperforming the repetition codes (or products thereof), one ought to increase the number of channel copies beyond $5$, resulting in code states on $10$ or more input qubits.
While the (polynomial) scaling of the neural network ansatz in the number of input qubits is favorable, the calculation of the coherent information is the bottleneck here: The computation for a code on $k$ qubits requires diagonalizing a dense $4^k\times 4^k$ matrix, which scales exponentially in runtime with the number of qubits.
Due to these computational limitations, evaluating the coherent information for $k\gtrsim7$ channel uses is thus an infeasible undertaking, and we would need to find an alternative approach---e.g.\ by exploiting symmetry considerations, or an approximate cost function that is faster to compute (see e.g.\ \cite{Wihler2014}, with the added difficulty that the coherent information is the difference between two entropies).

Furthermore, it could be possible that better quantum codes lie in maxima of measure almost zero, while the repetition code maxima dominate the potential landscape, making it difficult to find codes that surpass repetition codes.
In fact, in all our simulations for $k\leq 6$ copies of the depolarizing channel, the variational NN ansatz converges to product repetition codes.
Our results might be seen as indication that, among the states that can be represented using a neural network, repetition codes are in fact optimal for $k\leq 6$ copies of the depolarizing channel.
We note that our techniques of finding quantum codes using neural network states can also be applied to other channels such as generalized Pauli channels, which includes the depolarizing channel. 
A thorough investigation of other channels in this class, such as the BB84 channel \cite{BB84}, is the subject of ongoing work.

We also applied our ansatz to search for $\ame(n,d)$-states for values of $(n,d)$ for which it is unknown yet whether these states exist.
The smallest-dimensional instances of these cases are $(4,6)$ and $(7,4)$ (see \suppref{sec:ame}).
For $(n,d)=(4,6)$ the best value we obtained was $Q_{2}(\psi_{4,6}) \approx 0.9956$, which translates via \eqref{eq:trace-parameter-bound} to a bound on the average trace distance parameter of $D_2(\psi_{4,6})\lesssim 0.6429$. 
The state $\psi_{n,d}$ achieving these values is an RBM state with binary encoding and a hidden layer width of $M=12$.
For $(n,d)=(7,4)$, we obtained $Q_{3}(\psi_{7,4}) \approx 0.9962$, corresponding to $D_{3}(\psi_{7,4}) \lesssim 0.7870$, achieved by an FF state with binary encoding and hidden layers $(14,14,14)$ with activation functions $\cos$-$\relu$-$\relu$.
These results suggest that, assuming AME states do exist in these cases, one has to tweak the neural network ansatz or the numerical methods, or both, in order to obtain numerical instances of AME states.

\section*{Acknowledgments}
We thank Sergey Bravyi, Andr\'{a}s Gily\'{e}n, Ivan Glasser, Felix Huber, Graeme Smith, and Mark M.~Wilde for interesting discussions and helpful feedback.
J.\,B.\ is funded by the Draper's Company Research Fellowship at Pembroke College, Cambridge.
F.\,L.\ is supported by National Science Foundation (NSF) Grant No.~PHY 1734006.
The authors appreciate the hospitality during visits, provided by DAMTP at the University of Cambridge, IQIM at Caltech, JILA at the University of Colorado Boulder, and QuSoft at the University of Amsterdam, where parts of this work were completed.

\printbibliography[title={References},heading=bibintoc]

\appendix

\section{Codes for the Generalized Amplitude Damping Channel}\label{sec:gadc-codes}
In this section we provide an overview of the quantum codes for the GADC defined in \eqref{eq:gadc} found using the neural network state ansatz.
To benchmark these neural network quantum codes we use weighted repetition codes
\begin{align}
|\phi_k^\lambda\rangle \coloneqq \sqrt{\lambda} |0\rangle_R\ox |0\rangle^{\ox k}_{A} + \sqrt{1-\lambda} |1\rangle_R\ox |1\rangle_{A}^{\ox k},
\label{eq:weighted-repetition-code-app}
\end{align}
whose simple structure allows for an efficient computation of the coherent information $Q^{(1)}(\phi_k^\lambda,\cA_{\gamma,N}^{\ox k})$.
In the following, we first carry out this calculation, and then present the optimal neural network codes that we found for the GADC.

\subsection{Formula for the Coherent Information of Repetition Codes}

We first determine the action of the GADC $\cA_{\gamma,N} = \sum\nolimits_{i=1}^4 A_i\rho A_i^\dagger$ with $A_i$ as defined in \eqref{eq:gadc} on a single qubit:
\begin{align}
\begin{aligned}
\cA_{\gamma,N}(|0\rangle\langle 0|) &= (1-\gamma N ) |0\rangle\langle 0| + \gamma N |1\rangle\langle 1|\\
\cA_{\gamma,N}(|1\rangle\langle 1|) &= (\gamma - \gamma N) |0\rangle\langle 0| + (1-\gamma + \gamma N) |1\rangle\langle 1|\\
\cA_{\gamma,N}(|0\rangle\langle 1|) &= \sqrt{1-\gamma} |0\rangle\langle 1|\\
\cA_{\gamma,N}(|1\rangle\langle 0|) &= \sqrt{1-\gamma} |1\rangle\langle 0|.
\end{aligned}
\label{eq:gadc-qubit-action}
\end{align}
Setting $\sigma_{RB^k} = (\id_R\ox \cA_{\gamma,N}^{\ox k})(\phi_k^\lambda)$, we have
\begin{align}
\sigma_{B^k} &= \lambda \cA_{\gamma,N}(|0\rangle\langle 0|)^{\ox k} + (1-\lambda) \cA_{\gamma,N}(|1\rangle \langle 1|)^{\ox k},
\end{align}
which is a diagonal operator with eigenvalues 
\begin{align}
r_m\coloneqq \lambda (1-\gamma N)^{k-m} (\gamma N)^{m} + (1-\lambda) (\gamma-\gamma N)^{k-m} (1-\gamma + \gamma N)^{m}
\end{align}
with multiplicity $\binom{k}{m}$ for $m=0,\dots,k$.
Hence,
\begin{align}
S(B^k)_\sigma = - \sum_{m=0}^k \binom{k}{m} r_m \log r_m.\label{eq:gadc-rep-entropy-B}
\end{align}

For the state on the joint system, we have
\begin{align}
\sigma_{RB^k} &= \lambda |0\rangle\langle 0|_R \ox \cA_{\gamma,N}(|0\rangle\langle 0|)^{\ox k} + (1-\lambda) |1\rangle\langle 1|_R\ox \cA_{\gamma,N}(|1\rangle\langle 1|)^{\ox k}\\
&\quad {} + \sqrt{\lambda(1-\lambda)} \left[|0\rangle\langle 1|_R\ox \cA_{\gamma,N}(|0\rangle\langle 1|)^{\ox k} + |1\rangle\langle 0|_R \ox \cA_{\gamma,N}(|1\rangle\langle 0|)^{\ox k}\right]\\
&= \lambda |0\rangle\langle 0|_R \ox \left[(1-\gamma N ) |0\rangle\langle 0| + \gamma N |1\rangle\langle 1|\right]^{\ox k}\\
&\quad {} + (1-\lambda) |1\rangle\langle 1|_R \ox \left[ (\gamma - \gamma N) |0\rangle\langle 0| + (1-\gamma + \gamma N) |1\rangle\langle 1| \right]^{\ox k}\\
&\quad {} + \sqrt{\lambda(1-\lambda)}(1-\gamma)^{k/2} |0\rangle\langle 1|_R \ox |0\rangle\langle 1|^{\ox k}\\
&\quad {} + \sqrt{\lambda(1-\lambda)}(1-\gamma)^{k/2} |1\rangle\langle 0|_R \ox |1\rangle\langle 0|^{\ox k}.
\end{align}
This operator can be written as
\begin{align}
\sigma_{RB^k} &= \mu_0 \sigma^{(0)} + \sum_{\substack{s^k\in\lbrace 0,1\rbrace^k\colon \\ s^k\neq (0,\dots,0)}} \mu^{(0)}_{s^k} |0s^k\rangle\langle 0s^k| + \sum_{\substack{s^k\in\lbrace 0,1\rbrace^k\colon\\ s^k\neq (1,\dots,1)}} \mu^{(1)}_{s^k} |1s^k\rangle\langle 1s^k|,
\end{align}
where $\mu_0 = \lambda (1-\gamma N)^k + (1-\lambda)(1-\gamma + \gamma N)^k$,
\begin{align}
\sigma^{(0)} &= \frac{1}{\mu_0}\begin{pmatrix} \lambda (1-\gamma N)^k & \sqrt{\lambda(1-\lambda)}(1-\gamma)^{k/2}\\ \sqrt{\lambda(1-\lambda)}(1-\gamma)^{k/2} & (1-\lambda)(1-\gamma + \gamma N)^k \end{pmatrix}\label{eq:sigma0}
\intertext{in the basis $\lbrace |0\rangle_R|0\rangle_B^{\ox k}, |1\rangle_R|1\rangle_B^{\ox k}\rbrace$, and}
\mu^{(0)}_{s^k} &= \lambda (1-\gamma N)^{k-|s^k|} (\gamma N)^{|s^k|}\\
\mu^{(1)}_{s^k} &= (1-\lambda) (\gamma-\gamma N)^{k-|s^k|} (1-\gamma + \gamma N)^{|s^k|}.
\end{align}
Let $r$ denote one of the eigenvalues of the state $\sigma^{(0)}$ in \eqref{eq:sigma0}, let 
\begin{align}
s_m &\coloneqq \lambda (1-\gamma N)^{k-m} (\gamma N)^{m}\\
t_m &\coloneqq (1-\lambda) (\gamma-\gamma N)^{k-m} (1-\gamma + \gamma N)^{m}
\end{align}
for $m=1,\dots,k-1$, and 
\begin{align}
s_0 &= 0 & t_0 &= (1-\lambda)(\gamma-\gamma N)^k\\
s_k &= \lambda (\gamma N)^k & t_k &= 0.
\end{align}
Then the entropy of $\sigma_{RB^k}$ equals
\begin{align}
S(RB^k)_\sigma = \mu_0 h(r) - \mu_0 \log \mu_0 - \sum_{m=0}^k \binom{k}{m} (s_m\log s_m + t_m\log t_m).\label{eq:gadc-rep-entropy-RB}
\end{align}
The coherent information $Q^{(1)}(\phi_k^\lambda,\cA_{\gamma,N}^{\ox k}) = S(B^k)_\sigma - S(RB^k)_\sigma$ can now be efficiently computed using Eq.s~\ref{eq:gadc-rep-entropy-B} and \ref{eq:gadc-rep-entropy-RB} for blocklenghts up to $k=20$.

\subsection{Neural Network Codes for the GADC}
We list the best neural network codes found for the GADC $\cA_{\gamma,N}$ in the following tables:
\begin{itemize}
	\item Table \ref{tab:gadc-codes-N01}: $(\gamma,N) = (0.44035,0.1)$
	\item Table \ref{tab:gadc-codes-N02-app}: $(\gamma,N) = (0.41488,0.2)$
	\item Table \ref{tab:gadc-codes-N03-app}: $(\gamma,N) = (0.40102,0.3)$
	\item Table \ref{tab:gadc-codes-N04-app}: $(\gamma,N) = (0.39392,0.4)$
	\item Table \ref{tab:gadc-codes-N05-app}: $(\gamma,N) = (0.39169,0.5)$
\end{itemize}
A comparison of these codes to weighted repetition codes is plotted in Fig.~\ref{fig:gadc-codes} in the main text.

\begin{table}
	\centering
	\begin{tabular}{llll}
	\multicolumn{4}{c}{$(\gamma,N) = (0.41488,0.2)$}\\[0.25em]
	\toprule
	$|\nu_k\rangle$ & $s^n$ ($A^k|R$) & $\psi(s^n)$ & $\frac{1}{k}\Qone(\nu_k,\cA_{\gamma,N}^{\ox k})$\\
	\midrule 
	$k=3$ 
	& \phantom{$00$}$000|110$\phantom{$00$} & $-0.4861 - 0.3994i$& $1.6923\cdot 10^{-3}$\\
	& \phantom{$00$}$001|011$\phantom{$00$} & $-0.3394 - 0.2938i$&\\
	& \phantom{$00$}$010|011$\phantom{$00$} & $-0.3392 - 0.2937i$&\\
	& \phantom{$00$}$100|011$\phantom{$00$} & $-0.3393 - 0.2938i$&\\
	\midrule
	$k=4$ 
	& \phantom{0}$0110|1111$\phantom{0} & $+0.3918 + 0.0061i$& $1.4132 \cdot 10^{-3}$\\
	& \phantom{0}$1011|1111$\phantom{0} & $+0.3881 + 0.1248i$&\\
	& \phantom{0}$1101|1111$\phantom{0} & $+0.4014 + 0.1190i$&\\
	& \phantom{0}$1111|1000$\phantom{0} & $+0.7018 - 0.1096i$&\\ 
	\midrule
	$k=5$ 
	& $00110|01011$ &  $+0.4043 + 0.2010i$& $9.8025 \cdot 10^{-4}$\\
	& $10001|01011$ &  $+0.4485 + 0.0298i$&\\
	& $10111|11101$ &  $+0.7542 + 0.1591i$&\\
	\bottomrule
	\end{tabular}
	\caption{Table of the best neural network codes for the GADC $\cA_{\gamma,N}$ with $(\gamma,N) = (0.41488,0.2)$ and $k=3,4,5$ channel copies. Only the non-zero amplitudes $\psi(s^n)$ indexed by the basis string $s^n$ (with $n =2k$) are shown (see \eqref{eq:qudit-state-main}). The architecture used for the neural network codes is a feed-forward net with four hidden layers of width $2k$ each, activation functions $\cos$ and $3\times \tanh$, and a Cartesian output layer (see Sec.~\ref{sec:gadc}).}
	\label{tab:gadc-codes-N02-app}
\end{table}

\begin{table}
	\centering
	\begin{tabular}{llll}
		\multicolumn{4}{c}{$(\gamma,N) = (0.40102,0.3)$}\\[.25em]
		\toprule
		$|\nu_k\rangle$ & $s^n$ ($A^k|R$) & $\psi(s^n)$ & $\frac{1}{k}\Qone(\nu_k,\cA_{\gamma,N}^{\ox k})$\\
		\midrule 
		$k=3$ 
		& \phantom{$00$}$000|010$\phantom{$00$} & $+0.2566 - 0.3601i$& $2.1889\cdot 10^{-3}$\\
		& \phantom{$00$}$000|011$\phantom{$00$} & $+0.2802 - 0.3704i$&\\
		& \phantom{$00$}$001|100$\phantom{$00$} & $+0.2572 - 0.3607i$&\\
		& \phantom{$00$}$010|100$\phantom{$00$} & $+0.2572 - 0.3607i$&\\
		& \phantom{$00$}$100|100$\phantom{$00$} & $+0.2573 - 0.3607i$&\\
		\midrule
		$k=4$ 
		& \phantom{0}$0101|1110$\phantom{0} & $+0.1624 - 0.3629i$& $7.3635 \cdot 10^{-4}$\\
		& \phantom{0}$1010|1110$\phantom{0} & $+0.0641 - 0.5748i$&\\
		& \phantom{0}$1111|0001$\phantom{0} & $+0.0582 - 0.5819i$&\\
		& \phantom{0}$1111|1000$\phantom{0} & $+0.1858 - 0.3618i$&\\ 
		\bottomrule
	\end{tabular}
	\caption{Table of the best neural network codes for the GADC $\cA_{\gamma,N}$ with $(\gamma,N) = (0.40102,0.3)$ and $k=3,4$ channel copies. For details, see Tab.~\ref{tab:gadc-codes-N02-app}.
	For $k=5$ copies of the GADC $\cA_{\gamma,N}$ with $(\gamma,N) = (0.40102,0.3)$ the neural network ansatz did not find any codes with positive coherent information.}
	\label{tab:gadc-codes-N03-app}
\end{table}

\begin{table}
	\centering
	\begin{tabular}{llll}
		\multicolumn{4}{c}{$(\gamma,N) = (0.39392,0.4)$}\\[0.25em]
		\toprule
		$|\nu_k\rangle$ & $s^n$ ($A^k|R$) & $\psi(s^n)$ & $\frac{1}{k}\Qone(\nu_k,\cA_{\gamma,N}^{\ox k})$\\
		\midrule 
		$k=3$ 
		& \phantom{$00$}$000|000$\phantom{$00$} & $+0.3653 - 0.3328i$& $2.3456\cdot 10^{-3}$\\
		& \phantom{$00$}$000|010$\phantom{$00$} & $+0.3517 - 0.2356i$&\\
		& \phantom{$00$}$001|110$\phantom{$00$} & $+0.3498 - 0.2642i$&\\
		& \phantom{$00$}$010|110$\phantom{$00$} & $+0.3499 - 0.2661i$&\\
		& \phantom{$00$}$100|110$\phantom{$00$} & $+0.3558 - 0.2541i$&\\
		\midrule
		$k=4$ 
		& \phantom{0}$0100|1100$\phantom{0} & $-0.6091 + 0.2278i$& $1.7592 \cdot 10^{-3}$\\
		& \phantom{0}$0101|0010$\phantom{0} & $-0.1881 + 0.3964i$&\\
		& \phantom{0}$0110|0010$\phantom{0} & $-0.1877 + 0.3963i$&\\
		& \phantom{0}$1100|0010$\phantom{0} & $-0.1877 + 0.3963i$&\\
		\bottomrule
	\end{tabular}
	\caption{Table of the best neural network codes for the GADC $\cA_{\gamma,N}$ with $(\gamma,N) = (0.39392,0.4)$ and $k=3,4,5$ channel copies. For details, see Tab.~\ref{tab:gadc-codes-N02-app}..}
	\label{tab:gadc-codes-N04-app}
\end{table}

\begin{table}
	\centering
	\begin{tabular}{llll}
		\multicolumn{4}{c}{$(\gamma,N) = (0.39169,0.5)$}\\[0.25em]
		\toprule
		$|\nu_k\rangle$ & $s^n$ ($A^k|R$) & $\psi(s^n)$ & $\frac{1}{k}\Qone(\nu_k,\cA_{\gamma,N}^{\ox k})$\\
		\midrule 
		$k=3$ 
		& \phantom{$00$}$010|101$\phantom{$00$} & $+0.1894 + 0.3909i$& $2.3948\cdot 10^{-3}$\\
		& \phantom{$00$}$100|101$\phantom{$00$} & $+1.8943 + 0.3909i$&\\
		& \phantom{$00$}$110|011$\phantom{$00$} & $+0.1735 + 0.4322i$&\\
		& \phantom{$00$}$110|100$\phantom{$00$} & $+0.1735 + 0.4323i$&\\
		& \phantom{$00$}$111|101$\phantom{$00$} & $+0.1895 + 0.3909i$&\\
		\midrule
		$k=4$ 
		& \phantom{0}$0100|0111$\phantom{0} & $-0.2629 - 0.4120i$& $1.7913 \cdot 10^{-3}$\\
		& \phantom{0}$0100|1101$\phantom{0} & $-0.2287 - 0.3772i$&\\
		& \phantom{0}$0101|0110$\phantom{0} & $-0.2248 - 0.3721i$&\\
		& \phantom{0}$0110|0110$\phantom{0} & $-0.2258 - 0.3738i$&\\
		& \phantom{0}$1100|0110$\phantom{0} & $-0.2225 - 0.3708i$&\\
		\midrule
		$k=5$ 
		& $01100|01010$ &  $-0.3311 + 0.2694i$& $1.3393 \cdot 10^{-3}$\\
		& $10100|01010$ &  $-0.3337 + 0.3209i$&\\
		& $11000|01010$ &  $-0.3336 + 0.3208i$&\\
		& $11100|10110$ &  $-0.2911 + 0.3009i$&\\
		& $11100|11001$ &  $-0.3335 + 0.3208i$&\\
		\bottomrule
	\end{tabular}
	\caption{Table of the best neural network codes for the GADC $\cA_{\gamma,N}$ with $(\gamma,N) = (0.39169,0.5)$ and $k=3,4,5$ channel copies. For details, see Tab.~\ref{tab:gadc-codes-N02-app}.}
	\label{tab:gadc-codes-N05-app}
\end{table}

\section{Codes for the Dephrasure Channel}\label{sec:dephrasure-codes}
In the following, we give a summary of the results about the coherent information of the dephrasure channel $\cN_{p,q}$ (defined in \eqref{eq:dephrasure}) that were obtained in \cite{LLS18}.
These results are concerned with the one-way quantum capacity, as defined in Sec.~\ref{sec:capacity}; for a discussion of two-way capacities, see \cite{Pirandola2019}.

\subsection{Formula for the Coherent Information of Repetition Codes}

Superadditivity of the channel coherent information of the dephrasure channel can be achieved using a simple \emph{weighted repetition code}
\begin{align}
|\phi_k^\lambda\rangle \coloneqq \sqrt{\lambda} |0\rangle_R\ox |0^k\rangle_{A^k} + \sqrt{1-\lambda} |1\rangle_R\ox |1^k\rangle_{A^k},\label{eq:weighted-rep-code}
\end{align}
where $\lambda\in[0,1]$.
In \cite{LLS18}, the following formula is derived for its channel coherent information:
\begin{align}
\Qone(\phi_k^\lambda,\Npq^{\ox k}) = ((1-q)^k-q^k)h(\lambda) - (1-q)^k\left(1-u\artanh u  - \frac{1}{2}\log\left(1-u^2\right)\right),
\label{eq:coh-info-repetition-code}
\end{align}
where $h(\lambda) = -\lambda\log\lambda - (1-\lambda)\log(1-\lambda)$ is the binary entropy (in terms of the binary logarithm), $\artanh(x) \coloneqq \frac{1}{2}\log\frac{1+x}{1-x}$, and 
\begin{align}
u = u(\lambda,p,k) = \sqrt{1- 4\lambda(1-\lambda)(1-(1-2p)^{2k})}.
\end{align}
Moreover, it is shown in \cite{LLS18} that for $k=1$ the formula in \eqref{eq:coh-info-repetition-code} maximized over $\lambda\in[0,1]$ is in fact the optimal single-letter channel coherent information.
That is, $\Qone(\Npq)$ is optimized by states whose marginal on the system qubits is diagonal in the computational basis.
Hence, the formula \eqref{eq:coh-info-repetition-code} can be used to find quantum codes that surpass the optimal code for a single copy of $\Npq$, demonstrating superadditivity of coherent information.

For $q\in\lbrace 0.1,0.2,0.3,0.4\rbrace$ and the relevant intervals of $p$, the rates of the weighted repetition code $\phi_k^\lambda$ (optimized over $\lambda\in[0,1]$) for $1\leq k\leq 5$ are plotted in Fig.~\ref{fig:dephr-codes}.
The lines corresponding to $\phi_1$ represent the codes achieving the optimal single-letter coherent information $\Qone(\cN_{p,q})$.

\subsection{Neural Network Codes for the Dephrasure Channel}

We list the best neural network codes found for the dephrasure channel $\cN_{p,q}$ in the following tables:
\begin{itemize}
	\item Table \ref{tab:dephr-codes-q04}: $(p,q) = (0.08,0.4)$
	\item Table \ref{tab:dephr-codes-q03}: $(p,q) = (0.16,0.3)$
	\item Table \ref{tab:dephr-codes-q02}: $(p,q) = (0.24,0.2)$
	\item Table \ref{tab:dephr-codes-q01}: $(p,q) = (0.32,0.1)$
\end{itemize}
A comparison of these codes to weighted repetition codes is plotted in Fig.~\ref{fig:dephr-codes} in the main text.

\begin{table}
	\centering
	\begin{tabular}{llll}
		\toprule
		$|\nu_k\rangle$ & $s^n$ ($A^k|R$) & $\psi(s^n)$ & $\frac{1}{k}\Qone(\nu_k,\cN_{p,q}^{\ox k})$\\
		\midrule 
		$k=2$ 
		& \phantom{$00$}$00|11$\phantom{$00$} & $+0.1298-0.9905i$ & $2.1465\cdot 10^{-5}$\\
		& \phantom{$00$}$11|00$\phantom{$00$} & $+0.0255-0.0386i$&\\
		\midrule
		$k=3$ 
		& \phantom{0}$000|011$\phantom{0} & $+0.0008+0.0000i$& $3.9686 \cdot 10^{-5}$\\
		& \phantom{0}$001|100$ & $-0.6922-0.7114i$&\\
		& \phantom{0}$011|010$ & $-0.0005+0.0004i$&\\
		& \phantom{0}$1  0  1 | 0  1  1$& $-0.0007+0.0003i$ & \\
		& \phantom{0}$1  1  0 | 0  0  1$ & $+0.0754+0.0948i$ &\\
		& \phantom{0}$1  1  0 | 1  0  1$ & $-0.0005-0.0014i$ &\\
		\midrule
		$k=4$ 
		& $0  0  0  0 | 1  0  0  0$ & $+0.0941-0.1059i$ & $4.7922 \cdot 10^{-5}$\\
		& $0  0  1  0 | 1  0  0  0$ & $+0.0495+0.1017i$ &\\
		& $1  1  0  1 | 0  0  0  0$ & $+0.9362+0.3012i$ &\\
		\bottomrule
	\end{tabular}
	\caption{Table of the best neural network codes for the dephrasure channel $\cN_{p,q}$ with $(p,q) = (0.16,0.3)$ and $k=2,3,4$ channel copies. Only the non-zero amplitudes $\psi(s^n)$ indexed by the basis string $s^n$ (with $n =2k$) are shown (see \eqref{eq:qudit-state-main}). The architecture used for the neural network codes is a feed-forward net with four hidden layers of width $2k$ each, activation functions $\cos$ and $3\times \relu$, and a Polar output layer (see Sec.~\ref{sec:dephrasure-main}).}
	\label{tab:dephr-codes-q03}
\end{table}

\begin{table}
	\centering
	\begin{tabular}{llll}
		\toprule
		$|\nu_k\rangle$ & $s^n$ ($A^k|R$) & $\psi(s^n)$ & $\frac{1}{k}\Qone(\nu_k,\cN_{p,q}^{\ox k})$\\
		\midrule 
		$k=2$ 
		& \phantom{00} $0  1 | 0  0$ & $-0.0071+0.0005i$ & $6.8447 \cdot 10^{-6}$\\
		& \phantom{00} $0  1 | 0  1$ & $-0.0820+0.9959i$ &\\
		& \phantom{00} $1  0 | 1  0$ & $-0.0325+0.0202i$ &\\
		\midrule
		$k=3$ 
		& \phantom{0} $0  0  0 | 1  0  0$ & $-0.0004-0.0009i$ & $1.1382\cdot 10^{-5}$\\
		& \phantom{0} $0  0  1 | 0  0  0$ & $-0.0022+0.0048i$ &\\
		& \phantom{0} $0  0  1 | 0  0  1$ & $+0.0235+0.0011i$ &\\
		& \phantom{0} $0  0  1 | 0  1  0$ & $+0.0257-0.0019i$ &\\
		& \phantom{0} $0  0  1 | 1  0  1$ & $+0.1600+0.9795i$ &\\
		& \phantom{0} $0  0  1 | 1  1  0$ & $+0.0100+0.0094i$ &\\
		& \phantom{0} $0  1  1 | 1  1  0$ & $-0.0006-0.0009i$ &\\
		& \phantom{0} $1  0  1 | 1  1  0$ & $+0.0010-0.0005i$ &\\
		& \phantom{0} $1  1  0 | 0  1  1$ & $-0.0548-0.1031i$ &\\
		\midrule
		$k=4$ 
		& $ 0  1  1  0 | 1  1  1  0$ & $-0.1047+0.0727i$ & $1.1561\cdot 10^{-5}$\\
		& $ 0  1  1  1 | 1  1  1  0$ & $-0.0531+0.0863i$ &\\
		& $ 1  0  0  0 | 0  0  1  0$ & $+0.1405-0.9766i$ &\\
		& $ 1  0  0  1 | 1  1  0  0$ & $+0.0001+0.0000i$ &\\
		& $ 1  0  0  1 | 1  1  0  1$ & $+0.0003-0.0031i$ &\\
		\bottomrule
	\end{tabular}
	\caption{Table of the best neural network codes for the dephrasure channel $\cN_{p,q}$ with $(p,q) = (0.24,0.2)$ and $k=2,3,4$ channel copies. For details, see Tab.~\ref{tab:dephr-codes-q03}.}
	\label{tab:dephr-codes-q02}
\end{table}

\begin{table}
	\centering
	\begin{tabular}{llll}
		\toprule
		$|\nu_k\rangle$ & $s^n$ ($A^k|R$) & $\psi(s^n)$ & $\frac{1}{k}\Qone(\nu_k,\cN_{p,q}^{\ox k})$\\
		\midrule 
		$k=2$ 
		& \phantom{00} $0  0 | 1  1$ & $+0.2425-0.0318i$ & $9.9204\cdot 10^{-5}$ \\
		& \phantom{00} $0  1 | 0  0$ & $-0.0038-0.0074i$ & \\
		& \phantom{00} $1  0 | 0  0$ & $-0.0038-0.0074i$ &\\
		& \phantom{00} $1  1 | 0  0$ & $+0.0000-0.0001i$ & \\
		& \phantom{00} $1  1 | 0  1$ & $-0.3215-0.9147i$ & \\
		\midrule
		$k=3$ 
		& \phantom{0} $0  0  0 | 1  1  1$ & $-0.4863-0.8208i$ & $1.1172\cdot 10^{-4}$\\
		& \phantom{0} $0  1  0 | 1  0  1$ & $+0.0000-0.0001i$ & \\
		& \phantom{0} $0  1  1 | 0  0  0$ & $+0.0367-0.1624i$ & \\
		& \phantom{0} $1  0  0 | 0  0  0$ & $-0.0008+0.0009i$ & \\
		& \phantom{0} $1  0  1 | 0  0  0$ & $+0.0332-0.1628i$ & \\
		& \phantom{0} $1  1  0 | 0  0  0$ & $+0.0246-0.1643i$ & \\
		& \phantom{0} $1  1  1 | 0  0  0$ & $-0.0328-0.0762i$ & \\
		\midrule
		$k=4$ 
		& $0  0  0  0 | 1  1  0  1$ & $-0.1397-0.0564i$ & $1.1802\cdot 10^{-4}$\\
		& $0  1  0  0 | 0  0  1  0$ & $-0.0001-0.0004i$ & \\
		& $0  1  0  1 | 1  0  1  0$ & $-0.0171-0.0879i$ & \\
		& $0  1  1  0 | 0  1  1  0$ & $-0.5843+0.3576i$ & \\
		& $0  1  1  1 | 1  0  1  0$ & $+0.0002+0.0000i$ & \\
		& $1  0  0  1 | 0  0  0  0$ & $+0.0933-0.6787i$ & \\
		& $1  0  1  0 | 0  0  1  1$ & $-0.0001+0.0001i$ & \\
		& $1  0  1  0 | 0  1  1  1$ & $-0.0196-0.0874i$ & \\
		& $1  0  1  1 | 0  1  0  0$ & $+0.0004-0.0001i$ & \\
		& $1  1  1  1 | 0  0  1  1$ & $-0.0985-0.1140i$ & \\
		\bottomrule
	\end{tabular}
	\caption{Table of the best neural network codes for the dephrasure channel $\cN_{p,q}$ with $(p,q) = (0.32,0.1)$ and $k=2,3,4$ channel copies. For details, see Tab.~\ref{tab:dephr-codes-q03}.}
	\label{tab:dephr-codes-q01}
\end{table}

\section{Codes for the Depolarizing channel}\label{sec:depolarizing-codes}
\subsection{Product Repetition Codes for the Depolarizing Channel}\label{sec:product-rep-codes}
\begin{figure}[t]
	\centering
	\includegraphics{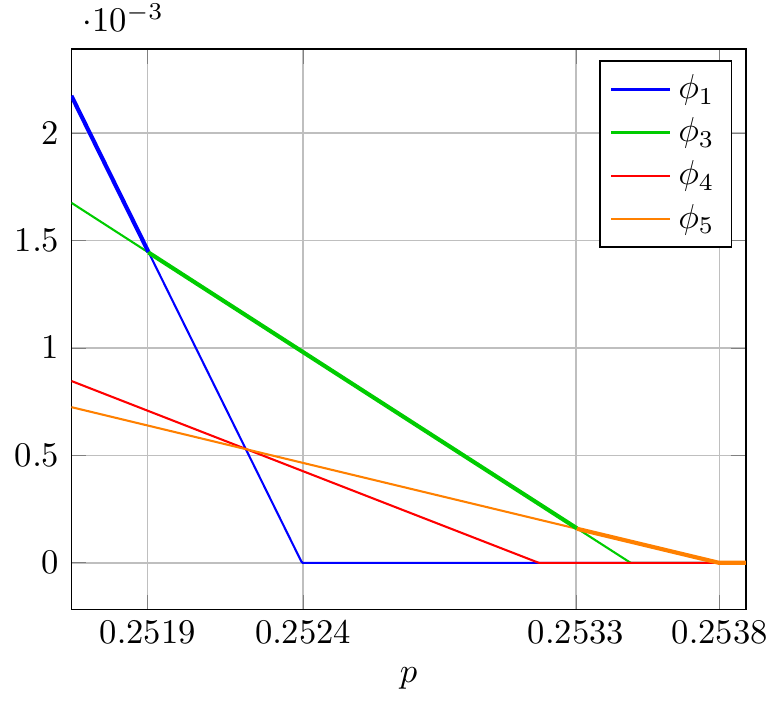}
	\caption{Rates and thresholds for the coherent information of repetition codes $\phi_k$ for the depolarizing channel $\cD_p^{\ox k}$ with $p\in [0.2516,0.2539]$ and $k=1,3,4,5$.
		The enveloping thick line marks the known optimal coherent information for the depolarizing channel (up to the concatenated codes of \cite{SS07,FW08}, which are not shown here).}
	\label{fig:optimal-repetition-codes}
\end{figure}

In this appendix, we discuss the known optimal codes for the depolarizing channel, which are given by repetition codes 
\begin{align}
|\phi_k\rangle=\frac{1}{\sqrt{2}}(|0\rangle_R|0\rangle_A^{\ox k} + |1\rangle_R|1\rangle_A^{\ox k}). \label{eq:repetition-code-app}
\end{align}
For $p\lesssim 0.2519$, the single-letter coherent information \eqref{eq:single-letter-cohinfo} is optimal.
For $0.2519\lesssim p\lesssim 0.2533$, the $3$-repetition code $\phi_3$ (defined in \eqref{eq:repetition-code-app}) is optimal, while for $0.2533\lesssim p \lesssim 0.2538$ the $5$-repetition code $\phi_5$ is optimal.
The point $p\simeq 0.2538$ marks the highest threshold for a single repetition code.
This threshold can be further extended using the concatenated codes of \cite{SS07,FW08}.\footnote{
	Note that these concatenated codes require at least $10$ channel uses of the depolarizing channel, and thus their rate is far lower than the rates of the codes just described.
	Furthermore, investigating $n\geq 10$ channel uses of the depolarizing channel is at the moment out of reach for our numerical methods.
	For these reasons, we focus on the regime $p\leq 0.2538$ within the threshold of $\phi_5$.}
We summarize this in Fig.~\ref{fig:optimal-repetition-codes}, where we compare the repetition codes and their rates and thresholds.

The above codes are the best known \emph{information-theoretic} codes, yielding the best lower bounds on the quantum capacity of the depolarizing channel by \eqref{eq:achievability}.
However, in numerical investigations we are facing a slightly different problem of maximizing the $k$-coherent information $\frac{1}{k} Q^{(1)}(\psi,\cD_p^{\ox k})$ over quantum codes $\psi$ for \emph{fixed} $k$, that is, solving
\begin{align}
\argmax_{\psi}\frac{1}{k} Q^{(1)}(\psi,\cD_p^{\ox k}).\label{eq:fixed-n-opt}
\end{align}

For $k\leq 9$ channel uses, the optimization problem \eqref{eq:fixed-n-opt} is solved by \emph{products} of repetition codes, 
\begin{align}
|\Phi_\mathbf{k}\rangle = \bigotimes_{i=1}^l|\phi_{k_i}\rangle.\label{eq:product-rep-codes}
\end{align}
Here, $\mathbf{k}=(k_1,\dots,k_l)$, and the resulting code $|\Phi_\mathbf{k}\rangle$ is a quantum code on $\sum_{i=1}^l k_i$ channel input qubits and $l$ purifying qubits.

\begin{figure}
	\centering
	\includegraphics[height=0.8\textheight]{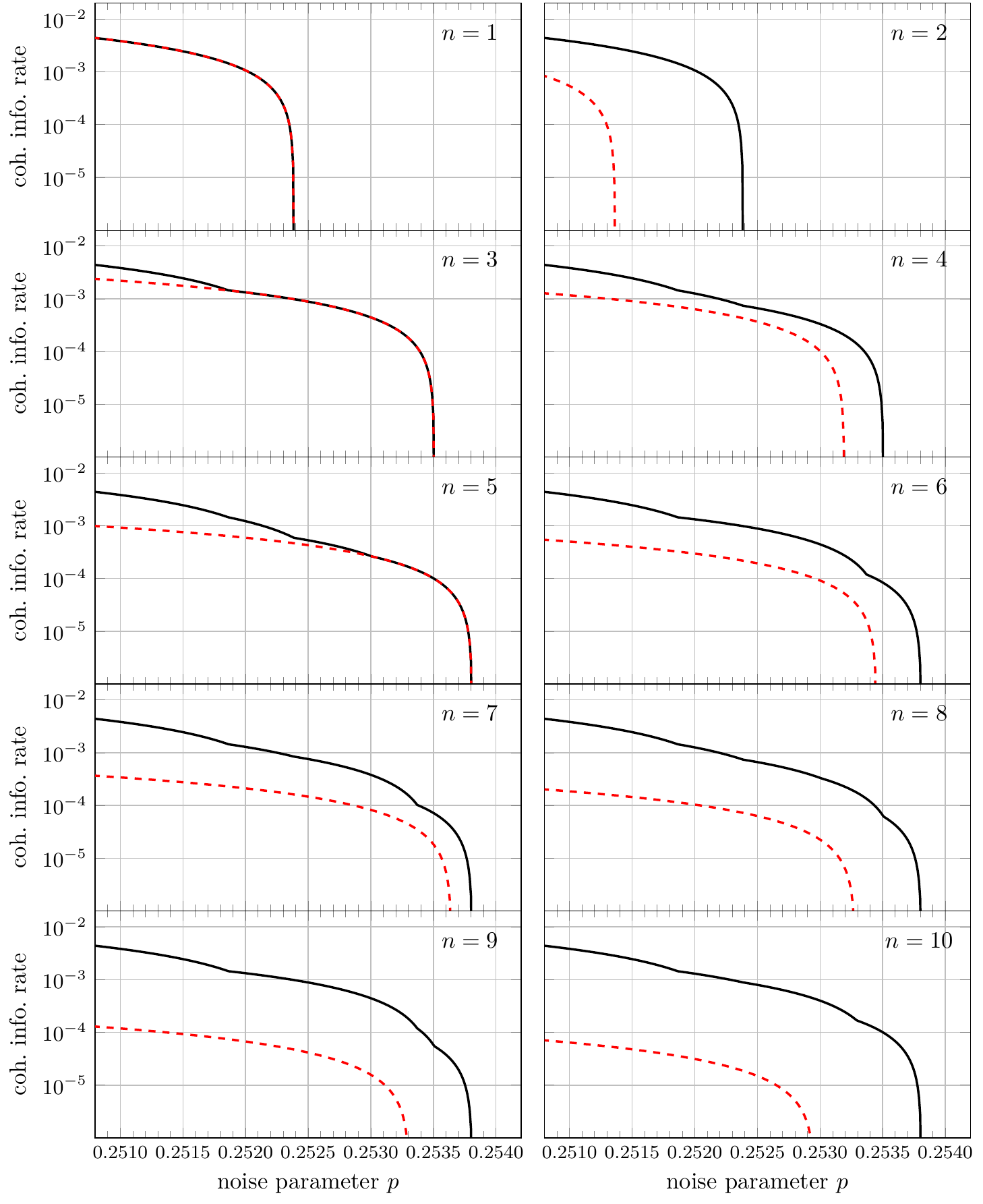}
	\caption{For $k=1,\ldots,10$ channel uses, the dashed red line is the coherent information rate of the $k$-repetition code.
		The solid black line is the best achievable rate when only using product codes, e.g.\ for $k=3$ and below $p\approx0.252$, a product of three single-channel repetition codes ($1\times1\times1$) is superior to one $3$-repetition code.
		It is noteworthy that the segmentation of the best achievable rates is not clear \emph{a priori}: For $k=4$, the segments are $1\times1\times1\times1$ and then $3\times1$, where the extra kink at $p\approx0.2524$ signifies that the single-letter CI has now dropped to zero;
		for $k=6$, the segments are $1\times\ldots\times1$, $3\times3$, and $5\times1$---the latter one of which is just a single segment, as the single-letter CI is already zero.}
	\label{fig:best-rep-codes}
\end{figure}

To illustrate this, consider 4 channel uses of the depolarizing channel, and recall that the single-letter coherent information \eqref{eq:single-letter-cohinfo} vanishes around $p\simeq 0.2524$.
The respective thresholds for the $4$-repetition code $\phi_4$ and the $3$-repetition code $\phi_3$ on \emph{three} input qubits are $p\simeq 0.2532$ and $p\simeq 0.2535$, respectively (see Fig.~\ref{fig:optimal-repetition-codes} and the file \texttt{rep-codes-tabular.txt} in \cite{anc_files}).
Hence, for $0.2532\leq p\leq 0.2535$ it is clearly advantageous to ``freeze'' one input qubit to some fixed pure state, and use a $3$-repetition code on the remaining $3$ input qubits.
Since pure input states can never establish coherent information between Alice and Bob, the frozen input does not contribute to the overall coherent information, and the resulting code incurs a penalty in the rate.
However, this code inherits the same threshold as the $3$-repetition code on three input qubits, thus outperforming the plain $4$-repetition code.
Similarly, one finds that for $p\in[0.2519,0.2524]$ the quantity $\frac{1}{4}Q^{(1)}(\cD_p^{\ox 4})$ is maximized by a $3$-repetition code tensored with a $1$-repetition code (i.e., using three of the four input qubits with one purifying qubit for a repetition code, and maximally entangling the remaining input qubit with another purifying qubit).
In Fig.~\ref{fig:best-rep-codes} and Tab.~\ref{tab:best-rep-codes-thresholds}, we provide an overview of the thresholds and rates of the optimal such combinations of repetition codes for $k\leq 10$ uses of the depolarizing channel.
For $k\geq 10$ uses of the depolarizing channel, concatenated codes can surpass the best known repetition code thresholds \cite{SS07,FW08}.

\begin{table}[h]
	\centering
	\renewcommand{\arraystretch}{1.2}
	\begin{tabular}{p{1cm}llllll}
		\hline
		&                                  \multicolumn{6}{c}{$p$}                                     \\
		\cline{2-7}
		$k$ & 0.25186               & 0.25238        & 0.25301 & 0.25329 & 0.25337        & 0.25350          \\ 
		\hline
		4   & $1\times3$               &                   &            &            &                   &  \\
		5   & $1\times1\times 3$       & $2\times3$        & $5\times 0$         &            &                   &  \\
		6   & $3\times3$               &                   &            &            & $1\times5$        &  \\
		7   & $1\times3\times3$        &                   &            &            & $2\times5$        &  \\
		8   & $1\times1\times3\times3$ & $2\times3\times3$ & $3\times5$ &            &                   &  \\
		9   & $3\times3\times3$        &                   &            &            & $1\times3\times5$ & $4\times5$   \\
		10  & $1\times3\times3\times3$ &                   &            & $5\times5$ &                   &              \\ 
		\hline
	\end{tabular}
	\caption{Intermediate product repetition code thresholds; before the first column at 0.25186 the best code is given by the single-letter coherent information.}
	\label{tab:best-rep-codes-thresholds}
\end{table}

\subsection{Products of Repetition Codes as Benchmark for Depolarizing Noise}\label{sec:analytical-proofs}
As a benchmark for finding quantum codes, we demand that the models we propose can at least achieve the product repetition codes described above;
either because they can represent products of repetition codes directly, or because they achieve the target rates by some other means.
In particular, this should serve as a sanity check for the models we propose, indicating whether we need to increase the width of a hidden layer, or the depth of the model.
The relevant question for us is whether a state $|\Phi_\mathbf{n}\rangle$ as defined in \eqref{eq:product-rep-codes} can always be represented accurately by the weights obtained from an RBM or an FF net.

\subsubsection{RBM States}
First observe how the Hamiltonian $\op H_\text{RBM}$ describes a linear single-layer FF classifier (i.e., a linear function on the inputs $i_k$).
Seen as a linear function on bit strings, the Hamiltonian can therefore represent a target state $\ket{\psi}$ as well as a linear model allows.
For the simple case of products of repetition codes, where we subdivide the set of basis states into those of weight 0 and 1, respectively, this question is well-studied in the context of linear classifiers.

A single $k$-repetition code has the form $\ket{0\cdots0} + \ket{1\cdots1}\eqqcolon\ket a+\ket b$.
Since the RBM uses a scaled encoding (see Tab.\ \ref{tab:encodings}), the bit strings correspond to real entries in a $k$-dimensional vector, and thus $\ket b = 0\ket a$;
a linear function $L$ therefore necessarily satisfies $L(\ket b) = L(\ket a) = 0$.
If we let $\ket b$ be a basis state (unnormalized) and complete the basis with $k-1$ arbitrary orthogonal vectors, it therefore suffices to define $L$ in such a way to have $\ker L=\lin\{ \ket b \}$.

Products of repetition codes always have the form $\bigotimes_{i=1}^k\ket{\phi_{n_i}}$; since basis states are bit strings for the RBM classifier, the corresponding code is a direct sum of the individual repetition codes.
We can thus construct a classifier for the overall code by writing $L_1\oplus\ldots\oplus L_k=L$, which is still linear.

Since $\op H_\text{RBM}$ appears in an exponential in \eqref{eq:rbm-state}, we can use the spectral gap of $\op H_\text{RBM}$ to obtain a lower bound on how close to zero an entry in the code can be set.
For instance, if we were to represent a 3-repetition code $\ket{000}+\ket{111}$, we can require that $\ket{111}$ is the eigenvector corresponding to the smallest eigenvalue of $\op H_\text{RBM}$;
all other binary strings should have an energy that is as large as possible, such that the exponential function suppresses the corresponding weight.
Consider the binary state $\ket{011}$, which has overlap $\sqrt{2/3}$ with $\ket{111}$ (assuming normalization).
If $\Delta$ is the spectral gap of $\op H_\text{RBM}$---i.e.\ the difference between the ground state energy and the second lowest eigenvalue---then $\bra{001}\op H_\text{RBM}\ket{001}=2\Delta/3$, yielding a lower bound between largest code weight and smallest code weight of $\exp(-\Delta)$.
To get an empirical estimate, assume we flip a single bit---e.g.\ $i_1$---in \eqref{eq:rbm-energy}.
How large can the energy difference be?
If all parameters are chosen (in magnitude) within a range $[-M,M]$, then a simple estimate would be $\Delta\le M+M^2$;
this is, of course, an upper bound to a lower bound.
In practice we found that $M=10$ is sufficient for our purposes.

\subsubsection{DBM States}
\eqref{eq:dbm-energy} introduces a quadratic term in the input.
Since one can easily embed a \textsc{1-in-3Sat} instance into a quadratic polynomial (for three boolean variables $v_1, v_2, v_3$ where \textsc{true}$=1$ and \textsc{false}=$0$ enforced by terms $v_i^2-v_i=0$, the equation $(v_1+v_2+v_3-1)^2=0$ if and only if exactly one of the $v_i$ is \textsc{true}; the existence or nonexistence of a root for the sum of all constraints thus answers the instance), it is clear that the discriminative power of DBM states should vastly outperform that of RBM states, albeit at a higher computational cost.
As discussed in the introduction, for various ground states of local Hamiltonians this intuition has empirically been shown to be correct.

\subsubsection{Feed Forward Network States}
It is easy to explicitly construct weights for an FF net that can represent any product repetition code.
As a first step, consider a single repetition code $\ket{\phi_n}$.
We set up a three-layer neural network from $n$ inputs, one hidden layer of width $1$, and a single output node (for simplicity we disregard the imaginary part for the state output in \eqref{eq:FF-architecture}).
The weights and activation functions to be chosen are
\begin{equation}\label{eq:rep-code-ff}
\{ x_i \}_{i=1}^{n+1} \longmapsto y:=\cos\left(\frac{2\pi}{n+1}\sum_{i=1}^{n+1} x_i\right) \longmapsto z:=\relu\left(\frac{y-\cos(2\pi/n+1)}{1-\cos({2\pi/n+1})}\right),
\end{equation}
and one can verify that the output is one on the all $1$s and $0$s input, and zero otherwise.
We refer to Sec.\ \ref{sec:activation-f} for a more detailed discussion.

For a product code given by some $\mathbf n=(n_1,\ldots,n_k)$, we simply partition the input nodes into $k$ subsets and dovetail those with a network given in \eqref{eq:rep-code-ff};
we obtain $k$ outputs $z_1,\ldots,z_k$.
Since we know that a logic AND gate corresponds to all the $z_i=1$, we can use a final $\relu(\sum_{i=1}^k z_i - k + 1)$ layer to enforce that the weights are $1$ if all individual segments are valid repetition codes, and $0$ otherwise---or merge all already existing $\relu$ nodes into one.
Observe that we could always implement a single $\cos$ node in this function with two $\relu$ nodes, followed by another $\relu$ node to combine the outputs (as in Sec.\ \ref{sec:activation-f}).
This would increase the hidden layer width by a factor of two; we can incorporate addition of the individual outputs into the last existing $\relu$ layer, so the depth should remain constant.

One immediate consequence is that any product code of $k$ repetition codes can always be represented by a network architecture where the first hidden layer has width $k$; and we in fact empirically found that the trained weights of the first layer are similar to those in \eqref{eq:rep-code-ff}.

A final note on the parameter range necessary for the argument:
the largest coefficients in absolute value in \eqref{eq:rep-code-ff} and its final AND node are $\max_i(1-\cos(2\pi/n(i)))^{-1}$, or $k-1$, whichever is larger.
Restricting the network's parameter range artificially below this threshold could result in worse representability of product repetition codes.

\subsubsection{Schmidt Network States}
The argument is similar as for feed-forward network states.
Note that, in general, Schmidt codes will be redundant, since for e.g.\ four channel uses we are forced to using more than just a single purifying qubit.
The fact that the neural net calculates Schmidt coefficients means that the repetition codes always uses as many purifying dimensions as system dimensions.

\section{Absolutely Maximally Entangled States}\label{sec:ame}
An $\ame(n,d)$-state is a pure state $|\psi_{n,d}\rangle \in (\mathbb{C}^d)^{\ox n}$ on $n$ qudits with local dimension $d\geq 2$ satisfying \begin{align}
\rho_\cS = \tr_{\cS^c}\psi_{n,d} = \frac{1}{|\cS|} I_\cS
\end{align} 
for every $\cS\subset[n]$ with $|\cS|=\lfloor \frac{n}{2}\rfloor$.
As mentioned in the main text, whether or not an $\ame(n,d)$-state exists depends on $n$ and $d$.
Using weight enumerator theory \cite{SL97,Rai98}, Scott proved that an $\ame(n,d)$-state can only exist if $n\leq 2(d^2-1)$ for even $n$, and $n\leq 2d(d+1)-1$ for odd $n$ \cite{Sco04}.
This technique was recently extended by Huber et al.~\cite{HESG18} to give further constraints on the existence of $\ame(n,d)$-states.
For fixed $n$ an $\ame(n,d)$-state always exists for sufficiently large local dimension $d$ \cite{HC13}.
For example, $\ame(n,d)$-states exist for $d$ a prime power and $n\leq d$ \cite{GBR04}.
Recently, it was proved in \cite{HOS17} that an AME state on seven qubits cannot exist.
This result completely settled the case of qubit AME states: they exist for $n=2,3,5,6$, and only for these $n$.
Furthermore, there are constructions for certain combinations of parameters $(n,d)$ \cite{GBR04,HC13,Hel13,GALRZ15,GRDZ17}.
We refer to Figure 2 in \cite{HESG18} as well as Problem 35 on the IQOQI Vienna Open Quantum Problems list \cite{opq} for a more complete overview of the known results about existence of $\ame(n,d)$-states.
Here, we merely mention that it is unknown whether $\ame(n,d)$-states exist for $(n,d) = (4,6)$ and $(n,d) = (7,4),(7,6)$.

Scott \cite{Sco04} proved that a multipartite state $|\psi_{n,d}\rangle\in (\mathbb{C}^d)^{\ox n}$ is AME if and only if the average linear entropy $Q_m(\psi_{n,d})=1$, where $Q_m(\cdot)$ is defined in \eqref{eq:Qm}.
Since we are searching for AME states by maximizing $Q_m(\cdot)$, we need to make sure that a state $\psi_{n,d}$ with $Q_m(\psi_{n,d})\approx 1$ is also approximately AME.
We determine the latter by introducing the average trace distance parameter $D_m$ defined in \eqref{eq:AME-dist} that measures the average trace distance between the marginals of $\psi_{n,d}$ on $m$ subsystems and the completely mixed state.
The average trace distance parameter $D_m(\cdot)$ can be bounded from above in terms of $Q_m(\cdot)$, as stated in \eqref{eq:trace-parameter-bound}.
We restate this bound here for the reader's convenience:
\begin{align}
D_m(\psi_{n,d}) \leq \sqrt{2 \log[d^m-(d^m-1)Q_m(\psi_{n,d})]}.\label{eq:Dm-bound}
\end{align}
To prove \eqref{eq:Dm-bound}, we use the quantum version of Pinsker's inequality, $
D(\rho\|\sigma) \geq \frac{1}{2}\|\rho-\sigma\|_1^2,$
where $D(\rho\|\sigma)=\tr(\rho\log\rho)-\tr(\rho\log\sigma)$ is the quantum relative entropy.
We also use the $2$-relative R\'{e}nyi entropy $D_2(\rho\|\sigma)=\log\tr (\rho^2 \sigma^{-1})$ \cite{Pet86}, and the well-known fact that $D(\rho\|\sigma) \leq D_2(\rho\|\sigma)$.

Observe first that, for $\pi=\frac{1}{d}I$, we have $D_2(\rho\|\pi) = \log \tr(\rho^2) + \log d$, and hence
\begin{align}
\tr(\rho^2) = \frac{1}{d} \exp(D_2(\rho\|\pi)) \geq \frac{1}{d} \exp(D(\rho\|\pi)) \geq \frac{1}{d} \exp\left(\frac{1}{2}\left\|\rho-\pi\right\|_1^2\right).
\end{align}
Abbreviating $\pi_\cS=\frac{1}{|\cS|}I_\cS$, we then bound
\begin{align}
Q_m(\psi_{n,d}) &= \binom{n}{m}^{-1} \frac{d^m}{d^m-1} \sum_{\cS\subset[n]\colon |\cS|=m} (1-\tr \rho_\cS^2)\\
&\leq \binom{n}{m}^{-1} \frac{d^m}{d^m-1} \sum_{\cS\subset[n]\colon |\cS|=m} \left(1- \frac{1}{d^m} \exp\left(\frac{1}{2}\left\|\rho_\cS-\pi_\cS\right\|_1^2\right) \right)\\
&= \frac{1}{d^m-1} \left( d^m - \binom{n}{m}^{-1} \sum_{\cS\subset[n]\colon |\cS|=m} \exp\left(\frac{1}{2}\left\|\rho_\cS-\pi_\cS\right\|_1^2\right) \right)\\
&\leq \frac{1}{d^m-1} \left( d^m - \exp\left(\frac{1}{2} D_m(\psi_{n,d})^2\right) \right),\label{eq:Q-bound}
\end{align}
where the last inequality follows from concavity of the function $x\mapsto -\exp(\frac{x^2}{2})$.
Rearranging \eqref{eq:Q-bound} yields \eqref{eq:Dm-bound}.

Since AME states are defined on tensor products of $d$-dimensional Hilbert spaces, the input string $i^n$ to the neural network computing the amplitude $\psi(i^n)$ in the ansatz \eqref{eq:qudit-state-main} is a $d$-ary string.
Depending on the local dimension, we use different encodings of this $d$-ary input string, as explained in App.~\ref{sec:encodings} below.

\section{Input Encoding of $d$-ary Strings for Neural Networks}\label{sec:encodings}
In order to parametrize quantum states on $n$ qubits, it is rather straightforward to use the neural network ansatz described in Sec.~\ref{sec:NN-states}.
In the case of $\ame(n,d)$-states with local dimension $d>2$, we slightly tweak the neural network ansatz.
To this end, we fix a basis $\lbrace |i\rangle\rbrace_{i=0}^{d-1}$ for $\mathbb{C}^d$, and express a general quantum state $\psi_{n,d}\in (\mathbb{C}^d)^{\ox n}$ as
\begin{align}
|\psi_{n,d}\rangle = \frac{1}{C} \sum_{i^n\in [d]_0^n} \psi(i^n) |i^n\rangle,\label{eq:qudit-state}
\end{align}
where $C$ is again a normalization constant ensuring $\langle \psi_{n,d}|\psi_{n,d}\rangle = 1$, and we use the notation $[d]_0\coloneqq \lbrace 0,\dots,d-1\rbrace$.
We consider three different options of encoding the $d$-ary input string $i^n$ in order to obtain the amplitudes $\psi(i^n)$ in \eqref{eq:qudit-state} from a neural network:
\begin{enumerate}
	\item\label{item:scaled} (Scaled) direct encoding: Use the $d$-ary string $i^n$ directly, with a possible scaling of the entries such that $i_k\in[0,1]$ for $k\in[n]$.
	
	\item\label{item:binary} Binary encoding: Convert each symbol $i_k\in[d]_0$ into a binary string, requiring $\lceil\log d\rceil$ `physical' qubits per 'logical' qudit of $\psi_{n,d}$, and use the resulting binary string of length $\lceil\log d\rceil n$ as the input to the neural network.
	
	Example: For $d=6$, the encoding is $0\mapsto 000$, $1\mapsto 111$, $\dots$, $5\mapsto 101$.
	
	\item\label{item:one-hot} One-hot encoding: Encode each symbol in a 'one-hot' vector of length $d$ and use the resulting binary string of length $dn$ as the input to the neural network.
	
	Example: For $d=6$, the encoding is $0\mapsto 000001$, $1\mapsto 000010$, $\dots$, $5\mapsto 100000$.
\end{enumerate}
We have found that the performance of the specific encoding used in the neural network optimization depends on the local dimension $d$.
For prime $d$, the neural network optimization using the scaled encoding converges quickly to known $\ame(n,d)$-states such as $\ame(4,7)$, as evident from Fig.~\ref{fig:ame-states} in the main text.
On the other hand, for composite $d$ the NN ansatz is more powerful using binary or one-hot encoding.
Since binary encoding has a smaller overhead in terms of the 'physical' qubits used in the ansatz ($\lceil\log d\rceil n$ vs.~$dn$), we use binary encoding for composite local dimension $d$.
We summarize the different encodings in Tab.~\ref{tab:encodings}.

\begin{table}
	\centering
	\renewcommand{\arraystretch}{1.2}
	\begin{tabular}{lll}
		\hline
		& encoding ($i\in [d]_0$) & \# input nodes\\
		\hline
		scaled & $i \mapsto i/(d-1)$ & $n$\\
		binary & $i \mapsto \bin_{\lceil \log d \rceil}(i)$ & $\lceil \log d \rceil n$\\
		one-hot & $i \mapsto e_i$ & $dn$\\
		\hline
	\end{tabular}
	\caption{Summary of the possible encodings of a symbol $i\in [d]_0$ in a $d$-ary string $i^n$ of length $n$.
		We denote by $\bin_k(m)$ the binary representation of $m$ of length $k$ (with leading zeros if necessary), and by $e_i$ a vector with a $1$ in the $i$-th component and $0$s elsewhere.}
	\label{tab:encodings}
\end{table}

\section{The Role of Activation Functions for Quantum Codes}\label{sec:activation-f}
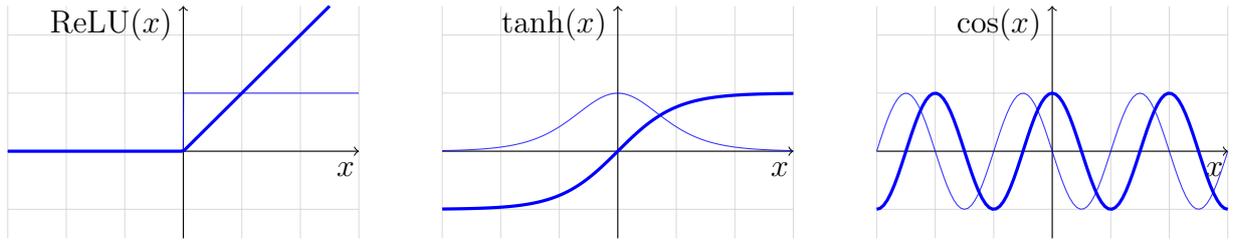
\begin{figure}[ht]
	\DeclareDocumentCommand\figureAAxis{m}{
		\draw[help lines, color=gray!30] (-3,-1.5) grid (3,2.5);
		\draw[->] (-3,0)--(3,0) node[below,xshift=-5]{$x$};
		\draw[->] (0,-1.5)--(0,2.5) node[left,yshift=-7]{$#1(x)$};
	}
	\begin{tikzpicture}[scale=.77]
	\figureAAxis{\relu}
	\draw[scale=1,domain=-3:2.5,samples=101,variable=\x,blue,very thick] plot ({\x},{max(\x,0)});
	\draw[scale=1,domain=-3:3,samples=1000,variable=\x,blue,opacity=.8] plot ({\x},{\x > 0});
	\end{tikzpicture}
	\hfill
	\begin{tikzpicture}[scale=.77]
	\figureAAxis{\tanh}
	\draw[scale=1,domain=-3:3,samples=101,variable=\x,blue,very thick] plot ({\x},{tanh(\x)});
	\draw[scale=1,domain=-3:3,samples=101,variable=\x,blue,opacity=.8] plot ({\x},{pow(1/cosh(\x),2)});
	\end{tikzpicture}
	\hfill
	\begin{tikzpicture}[scale=.77]
	\figureAAxis{\cos}
	\draw[scale=1,domain=-3:3,samples=101,variable=\x,blue,very thick] plot ({\x},{cos(180*\x)});
	\draw[scale=1,domain=-3:3,samples=101,variable=\x,blue,opacity=.8] plot ({\x},{-sin(180*\x)});
	\end{tikzpicture}
	\caption{Various activation functions (bold lines) and their derivatives (thin lines). $\tanh$ is an example for a sigmoid function; more commonly used, however, is $\sigmoid(x)=(1+\exp(-x))^{-1}$.
		It is clear that sigmoid functions suffer from a vanishing gradient problem on both ends of its input.
		This can be countered either by going to another activation function---such as a rectified linear unit $\relu$ (or its ``leaky'' version, i.e.\ one where the segment for $x<0$ has a small but non-vanishing slope), or using techniques such as batch normalisation \cite{Ioffe2015}.
		Non-monotonic activation functions such as $\cos$ are rarely used in practice, but can be useful for certain specific tasks.}
	\label{fig:activation-f}
\end{figure}

In machine learning, the use of nonlinear activation functions is crucial to a neural network's performance; otherwise, the network is just a single affine transformation and not useful beyond linear regression.
The overall network can have varying activation functions per neuron (see Fig.\ \ref{fig:networks}).
In essentially all cases, the activation functions are the same within a layer.
The operation of such a layer is thus to perform an affine transformation on the input vector and then, element-wise, apply the nonlinearity $f$. 
For a single neuron $z$ depending on $x=(x_1,\dots,x_n)$, the mathematical operation can thus be visualized as
\begin{align*}
\begin{tikzpicture}[
neuron/.style={
	circle, draw=black, fill=white, minimum size=.7cm, inner sep=0pt
}
]
\foreach\y/\l in {1/1, 2/2, 3/3, 5/n} {
	\draw (0, \y) node[left] {$x_\y$} -- (1.3, \y) to[out=0, in=180] (5, 3) node[pos=.7,neuron] {$w_{\y}$};
};
\node[] at (.91, 4.1) {$\vdots$};
\draw (5, 3) node[neuron,xshift=-10] {$\sum$} -- (7, 3) node[neuron] {$f$} -- (8, 3) node[right] {$z=f(\sum_iw_i x_i + b)$.};
\end{tikzpicture}
\end{align*}
Commonly used activation functions are e.g.\ $\relu$, $\sigmoid$ or $\tanh$, which are plotted in Fig.\ \ref{fig:activation-f};
in addition to some thorough studies \cite{Ioffe2015,He2015a,Krizhevsky2017}, there seems to be a lot of empirical understanding which activation functions perform better in various scenarios \cite{ListOfActivationFunctions}.
One example is that e.g.\ $\sigmoid$ saturates (meaning the gradient vanishes for large or small values), whereas e.g.\ $\relu$ does not have the same problem.
Furthermore, the general consensus seems to be that non-monotonic or periodic activation functions---such as e.g.\ $\sin$---weaken the neural network's performance.
We found conflicting evidence for this in the literature (\cite{Sopena1999,Gashler2016} and \cite[sec.~6.2.2]{Goodfellow2016}),
suggesting that such periodic functions can indeed be useful for specific tasks---especially in the context of representing ground states for local Hamiltonians \cite{CL18}.

In one example of such a task, \cite{CL18} use neural network states to approximate the ground states of certain Hamiltonians.
They report good performance of feed-forward network architectures with a cosine activation function in the first layer for a 1D anti-ferromagnetic Heisenberg model, arguing that the cosine function is capable of handling the ``sign problem'' typically found in the analysis of Hamiltonians.
We found that using cosine in the first hidden layer also performs well in finding good quantum codes for quantum channels such as the depolarizing channel defined in \eqref{eq:depolarizing}, or the dephrasure channel defined in \eqref{eq:dephrasure}.

In the following, we want to give an intuition why a periodic activation function such as $\cos$ can be useful for learning quantum codes with a structure that can be easily derived from the binary signature of its state vector.
To give an example, consider a repetition code on five qubits, given by $\ket{00000} + \ket{11111}$.
A function $M\colon(\field C^2)^{\otimes 5}\rightarrow\field C$ with $M(\ket{00000})=M(\ket{11111})=1$, and $0$ elsewhere, is trivial to construct from elementary logic gates (i.e.\ either all bits are zero, or all bits are one).

For a feed-forward neural network, one could imagine adding up all bits within one neuron, and thresholding this value with a $\relu$ activator:
\[
z_1 = \relu\left(2\sum_{i=1}^5 x_i - 9\right) = \begin{cases}
1 & \text{if $x_i=1\ \forall i$} \\
0   & \text{otherwise.}
\end{cases}
\]
A similar gate with flipped signs can activate only when all bits are zero; the two outputs can then be combined using a final $\relu$ node.

We can achieve the same activation using a single $\cos$ neuron, dovetailed by a $\relu$ in the next layer:
\[
z_1 = \cos\left( \frac{2\pi}{5}\sum_{i=1}^5 x_i \right)
\quad\text{and}\quad
z_2 = \relu\left( \frac{z_1-\cos(1/5)}{1-\cos(1/5)} \right).
\]
While this looks like a more complicated version of the same calculation,
it quickly becomes obvious that one can easily perform modular arithmetic using this technique---what we have in fact calculated is whether $\sum_ix_i\equiv 0\pmod 5$.

Why is this an advantage?
As a slightly more complicated example, let us consider an (unnormalized) tensor code built from a 3-repetition code $|\phi_3\rangle = |000\,000\rangle + |111\,111\rangle$ and a 1-repetition code (or simply maximally entangled state) $|\phi_1\rangle = |0\,0\rangle + |1\,1\rangle$.
In both cases, the first block of qubits (3 resp.~1) is sent through the channel, and the second block form the purifying environment.
On 4 qubits, the tensor code thus looks as follows (for visualization purposes we boldface the single channel repetition code):
\begin{align}\label{eq:3x1-channel-rep}
|\phi_3\rangle\ox |\phi_1\rangle = 	\ket{000\mathbf{0}\,000\mathbf{0}} + \ket{000\mathbf{1}\,000\mathbf{1}} + \ket{111\mathbf{0}\,111\mathbf{0}} + \ket{111\mathbf{1}\,111\mathbf{1}}
\end{align}
Any tensor channel $\cN^{\ox n}$ is naturally covariant\footnote{A quantum channel $\cM\colon A\to B$ is \emph{covariant} with respect to a group $G$ if there are unitary representations $g\mapsto U_A(g)$ on $\cH_A$ and $g\mapsto U_B(g)$ on $\cH_B$ such that $ \cM( U_A(g)\cdot U_A(g)^\dagger) = U_B(g) \cM(\cdot) U_B(g)^\dagger$ for all $g\in G$.} with respect to permuting tensor factors, i.e., the unitary representation $\pi\mapsto U_\pi$ of the symmetric group $S_n$ on $(\mathbb{C}^2)^{\otimes n}$ defined by $U_\pi |e_1\rangle\ox\dots\ox|e_n\rangle = |e_{\pi^{-1}(1)}\rangle\ox \dots\ox |e_{\pi^{-1}(n)}\rangle$.
Since the coherent information $I(A\rangle B)$ is furthermore invariant under local unitaries of the form $U_A\ox U_B$, codes that are permutations of each other yield the same value for the coherent information.
For example, the code
\begin{align}
(U_{(14)} \otimes U_{(24)}) (|\phi_3\rangle\ox |\phi_1\rangle) = \ket{\mathbf{0}000\,0\mathbf{0}00} + \ket{\mathbf{1}000\,0\mathbf{1}00} + \ket{\mathbf{0}111\,1\mathbf{0}11} + \ket{\mathbf{1}111\,1\mathbf{1}11}\label{eq:permuted-code}
\end{align}
is obtained from $|\phi_3\rangle\ox |\phi_1\rangle$ by swapping channel qubits 1 and 4 and environment qubits 2 and 4,\footnote{Note that in \eqref{eq:permuted-code} the two tensor products on the left-hand side are with respect to different tensor factors. For the first tensor product, the two factors correspond to channel input and purifying qubits, respectively.} and is thus equivalent for quantum information transmission.\footnote{We do \emph{not} claim that optimal codes are in any way symmetric due to this permutation invariance.}
Hence, within each block of four qubits (either channel or environment) the code is characterized by the Hamming weight of the code vectors (0, 1, 3 and 4 in the example above), and ideally this is identified by the neural network.
With modular arithmetic, we can have a $\cos$ neuron identifying 0 and 4 (e.g.\ all Hamming weights $\equiv 0\pmod 4$), and another one identifying 1 and 3 (e.g.\ all odd Hamming weights).

While it is conceivable that for simple codes such as \eqref{eq:3x1-channel-rep} one can write down relatively simple circuits with non-periodic activation functions, it should be clear that we do save space within the neural network representation if we can perform calculations such as the ones above within a single neuron.

\section{Numerical Optimization Techniques}\label{sec:numerics}
In most applications neural networks are trained using the backpropagation method, in which each network parameter is updated using the gradient of a loss or objective function with respect to that parameter.
In our main application of neural networks, maximizing the coherent information of a quantum channel, the objective function is the coherent information itself.
In the interesting case of a high-noise quantum channel (such as $\cD_p$ for $p\gtrsim 0.2523$), a randomly selected quantum code (e.g., with respect to the Haar measure on pure states) has strictly negative coherent information with high probability, whereas a product state $|\psi_1\rangle_R\ox|\psi_2\rangle_A$ always has vanishing coherent information, $I_c(\psi_1\ox\psi_2,\cN) = S(\cN(\psi_2)) - S(\psi_1\ox\cN(\psi_2)) = 0$.
Hence, the coherent information landscape is dominated by local maxima, and gradient-based optimization techniques are likely to get stuck in these local maxima.

This intuition was confirmed in our numerical search for good quantum codes for the depolarizing channel and the dephrasure channel.
In the search for $\ame(n,d)$ states, the objective function is the function $Q_m(\psi)$ defined in \eqref{eq:Qm}.
Here, numerical investigations also showed that gradient-based optimization was again likely to get stuck in local minima.

The failure of gradient-based optimization methods in both scenarios led us to consider gradient-free, stochastic global optimization techniques instead.
In the following, we give high-level explanations of four popular such algorithms, particle swarm optimization, artificial bee colonization, pattern search (also known as direct search), and genetic evolution.

\subsection{Particle Swarm Optimization}
\emph{Particle swarm optimization} (PSO) \cite{KE95} is a meta-heuristic, derivative-free global optimization technique.
The idea of PSO is to have multiple \emph{particles} explore the landscape on the search for a global minimum, and communicate their individual best value to the swarm. 
At the same time, each particle records its own history and stores the personal best value.
In each iteration, the update of a particle's velocity vector is determined by the current velocity, recurrence to the location of the personal best function value, and attraction towards the location of the global best value.

More precisely, fix model parameters $\alpha,\beta,\gamma >0$ and consider $N$ particles with random initial position $\mathbf{x}_i^{(0)}$ and random initial velocity $\mathbf{v}_i^{(0)}$ for $i\in[N]$.
For each particle $i$, the variable $\mathbf{p}_i$ stores the location of the personal best function value, while the variable $\mathbf{g}$ stores the location of the global best function value among the whole swarm.
In the $k$-th iteration, the velocity and position of a particle are updated according to
\begin{align}
\mathbf{v}_i^{(k)} &= \alpha \mathbf{v}_i^{(k-1)} + \beta r_\beta \left(\mathbf{p}_i - \mathbf{x}_i^{(k-1)} \right) + \gamma r_\gamma \left( \mathbf{g}-\mathbf{x}_i^{(k-1)}\right) \\
\mathbf{x}_i^{(k)} &= \mathbf{x}_i^{(k-1)} + \mathbf{v}_i^{(k)},
\end{align}
where $r_\beta,r_\gamma\in[0,1]$ are drawn uniformly at random.
The parameter $\alpha$ is called \emph{inertia}, while $\beta$ and $\gamma$ are usually called \emph{self-interaction} and \emph{social interaction}, respectively.
A common modification of the particle swarm optimization is to limit the social interaction to neighborhoods of a certain size within the swarm, ensuring a more thorough exploration of the landscape by the swarm.

The MATLAB implementation of PSO, available in the Global Optimization Toolbox, uses the neighborhood modifications with variable neighborhood sizes and an adaptive adjustment of the inertia weight.
We refer to the official documentation \cite{pso_matlab} for details of the algorithm, as well as the MATLAB files in \cite{anc_files} for the algorithm settings used in this paper.
Furthermore, we used the ``inertia weight'' variant of PSO in Pagmo \cite{abc_pagmo}, with parameter settings as found in the C++ source files \cite{anc_files}.

\subsection{Artificial Bee Colonization}
\emph{Artificial bee colonization} (ABC) \cite{Kar05} is another meta-heuristic, derivative-free global optimization technique based on the principle of swarm intelligence.
The algorithm works as follows: The population consists of $N$\emph{employer bees} and $N$ \emph{onlooker bees}.
While the employer bees explore the neighborhood of randomly created `food sources' (i.e., points in the landscape with a low objective function value for a minimization problem), the onlooker bees evaluate the food sources according to the promise given by the \emph{fitness} of the food source, and join the employer bees in exploring the neighborhood of those food sources.
If an employer bee cannot find any new food around its location for a certain number of iterations (i.e., it fails to find points in the neighborhood of the food source with a lower objective function value), it is converted into a \emph{scout bee} and assigned to a new random food source.

In more detail, to minimize a function $f\colon \mathbb{R}^D\to\mathbb{R}$, an employer bee at site $x_i$ randomly explores the neighborhood of $x_i$ by probing the location $x_i'$ which differs from $x_i$ in exactly one randomly drawn component $j\in[D]$ according to
\begin{align}
(x'_i)_j &= (x_i)_j + r((x_i)_j-(x_k)_j),
\end{align}
where $x_k\neq x_i$ is another randomly drawn food source, and $r\in[-1,1]$ is a uniform random number.
If $f(x_i')<f(x_i)$, the employer bee switches to $x'_i$ and continues exploring its neighborhood.
The fitness of the food source $x_i$ is defined as $\fit_i \coloneqq (1+f(x_i))^{-1}$, and each onlooker bee reinforces the employer bee group by selecting a food source according to the probability distribution $\lbrace \fit_i/\sum_i \fit_i\rbrace_i$.

We use the standard implementation of ABC found in the C++ optimization library Pagmo \cite{abc_pagmo}, as well as our own implementation of the standard algorithm in MATLAB (see \cite{anc_files}).

\subsection{Pattern Search}\label{sec:pattern-search} 
The third derivative-free optimization technique we use in this paper is called \emph{pattern search} or \emph{direct search}.
To minimize a function $f\colon \mathbb{R}^D\to \mathbb{R}$, the algorithm takes as input a starting point $x_0\in\mathbb{R}^D$ together with the objective function value $f(x_0)$, and creates a mesh of probing points around the starting point. 
In each iteration or \emph{poll}, the objective function is evaluated at each mesh point.
If for one of the mesh points, say $x_1$, the objective function value is lower than the current one (at $x_0$), the algorithm centers at $x_1$ and creates a new mesh.

There are different ways in how the mesh at a new center point is created. 
In a popular variant called \emph{generalized pattern search} (GPS), the new probing points $y_i$ of the mesh are defined by a fixed set $\cS\subset\mathbb{R}^D$ of vectors.
Common choices are $\cS_{2D}=\lbrace \pm\mathbf{e}_i\rbrace_{i=1}^D$, where $\mathbf{e}_i$ denotes the $i$-th standard basis vector, or $\cS_{D+1}=\lbrace \mathbf{e}_i\rbrace_{i=1}^D\cup \lbrace -(\mathbf{e}_1 + \dots + \mathbf{e}_D)\rbrace$.
In the $k$-th round with center point $x_{k-1}$, the points of the mesh are defined as $y_i= x_{k-1} + \Delta v_i$, where $v_i\in\cS$, and $\Delta$ is the \emph{mesh constant}. 
In a successful poll (i.e., when a new point with a lower objective function value is found), the mesh constant for the new mesh is doubled.
If the poll is unsuccessful, the center point remains the same and $\Delta$ is halved.

Another popular variant is called \emph{mesh adaptive direct search} (MADS).
Here, the set $\cR\subset\mathbb{R}^D$ of vectors for the new mesh points is randomly created after each successful poll.
In analogy to the GPS variant above, common choices are $\cR_{2D} = \lbrace \pm \mathbf{v}_i\rbrace_{i=1}^D$ and $\cR_{D+1} = \lbrace \mathbf{v}_i\rbrace_{i=1}^D \cup \lbrace -(\mathbf{v}_1 + \dots + \mathbf{v}_D)\rbrace$, where in each case the $\mathbf{v}_i$ are random vectors.

The above variants of pattern search are available in the Global Optimization Toolbox of MATLAB \cite{ps_matlab}.
We refer to the MATLAB files in \cite{anc_files} for the algorithm settings used in this paper.

\subsection{Simple Genetic Algorithm}\label{sec:sge}
The fourth derivative-free optimization algorithm is a genetic algorithm, which is related to evolutionary methods such as PSO and ABC, but motivated from the process of gene evolution.

Starting from a random selection of $N$ so-called ``chromosomes'' $\mathbf x_i^{(0)}$---where each vector component is called a ``gene''---a traditional implementation follows four steps.
\begin{itemize}
\item[Selection.] Pick random tuples of size $s$ from the chromosome pool, and select the ones with the best function value within each tuple; this creates a selected chromosome pool of size less than $N$.
\item[Crossover.] Randomly select a parent tuple (can be more than two, and up to the entire selected pool).
Merge the parents, e.g.\ by selecting a random chromosome, and replacing each gene (coordinate of $\mathbf x_i^{(0)}$) with some probability $p$ by genes from other chromosomes.
Continue creating child chromosomes until the new pool reaches size $N$.
\item[Mutation.] Randomize child genes within each chromosome according to some randomness distribution $\mathcal D$ and mutation probability $m$;
a popular variant of which is called \emph{polynomial mutation} where $\mathcal D\sim1/\poly$, which introduces a stronger bias towards creating children close to their parents.
\item[Reinsertion.] Merge parent and child chromosome pool and select $N$ of the fittest candidates.
\end{itemize}
We use Pagmo's standard implementation of a simple genetic algorithm with polynomial mutation (SGE, \cite{sge_pagmo}), with parameters $s=2$, $p=0.9$ and $m=0.02$.

\section{Additional Numerical Data}\label{sec:extra-numerics}
\newcommand{\extrafig}[4]{
\begin{figure}[h!]
	\includegraphics[width=\textwidth]{scatter-#1.pdf}
	\caption{Training convergence of #2, maximizing the CI of #3 copies of the depolarizing channel $\cD_p$, with noise parameter $p=0.2523$.
	The network architectures are identical to the ones in Fig.\ \ref{fig:convergence-dep}. #4}
	\label{fig:additional-#1}
\end{figure}
}
\extrafig{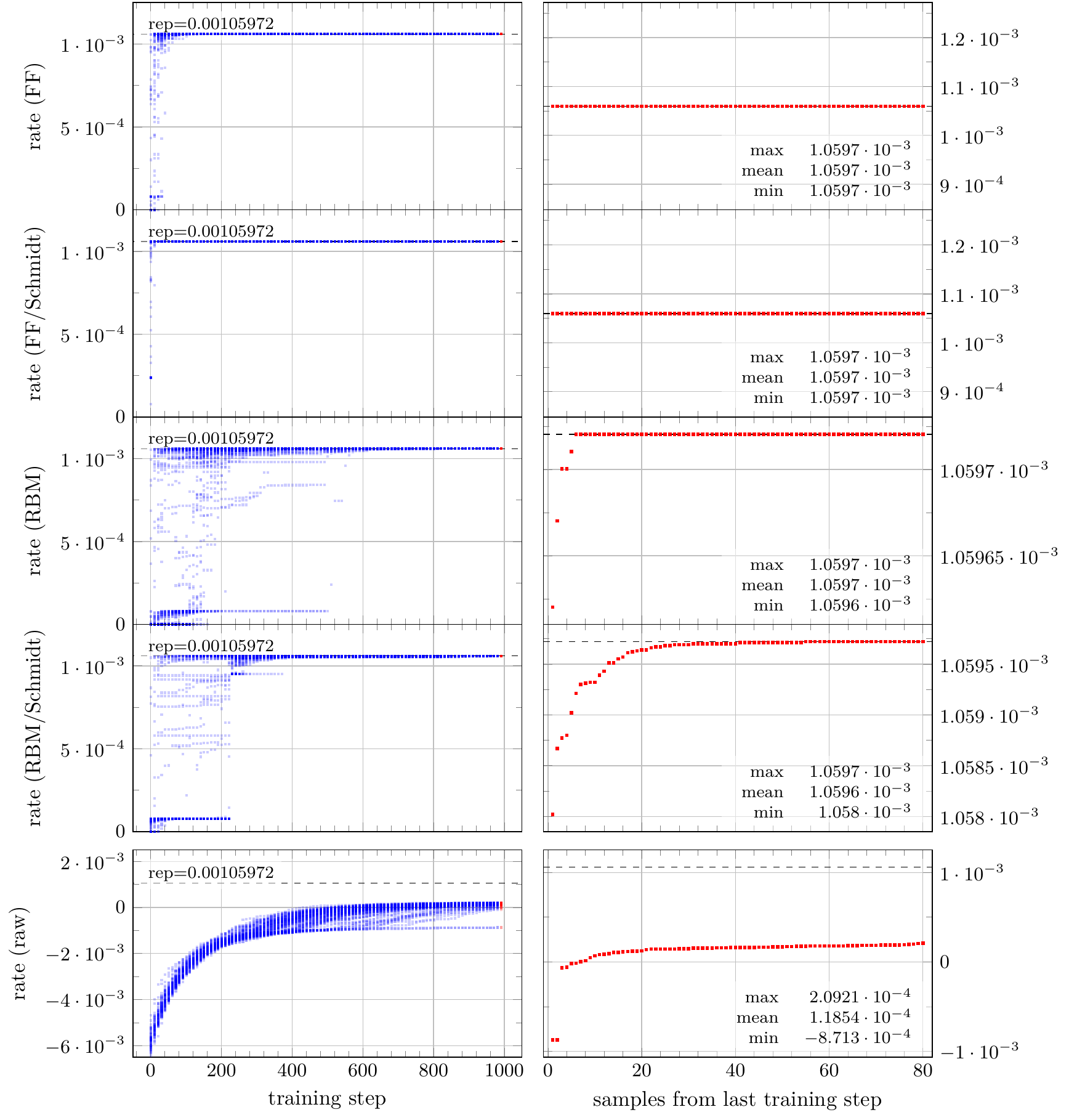}{an artificial bee colony (ABC) algorithm implemented in pagmo}{three}{}
\extrafig{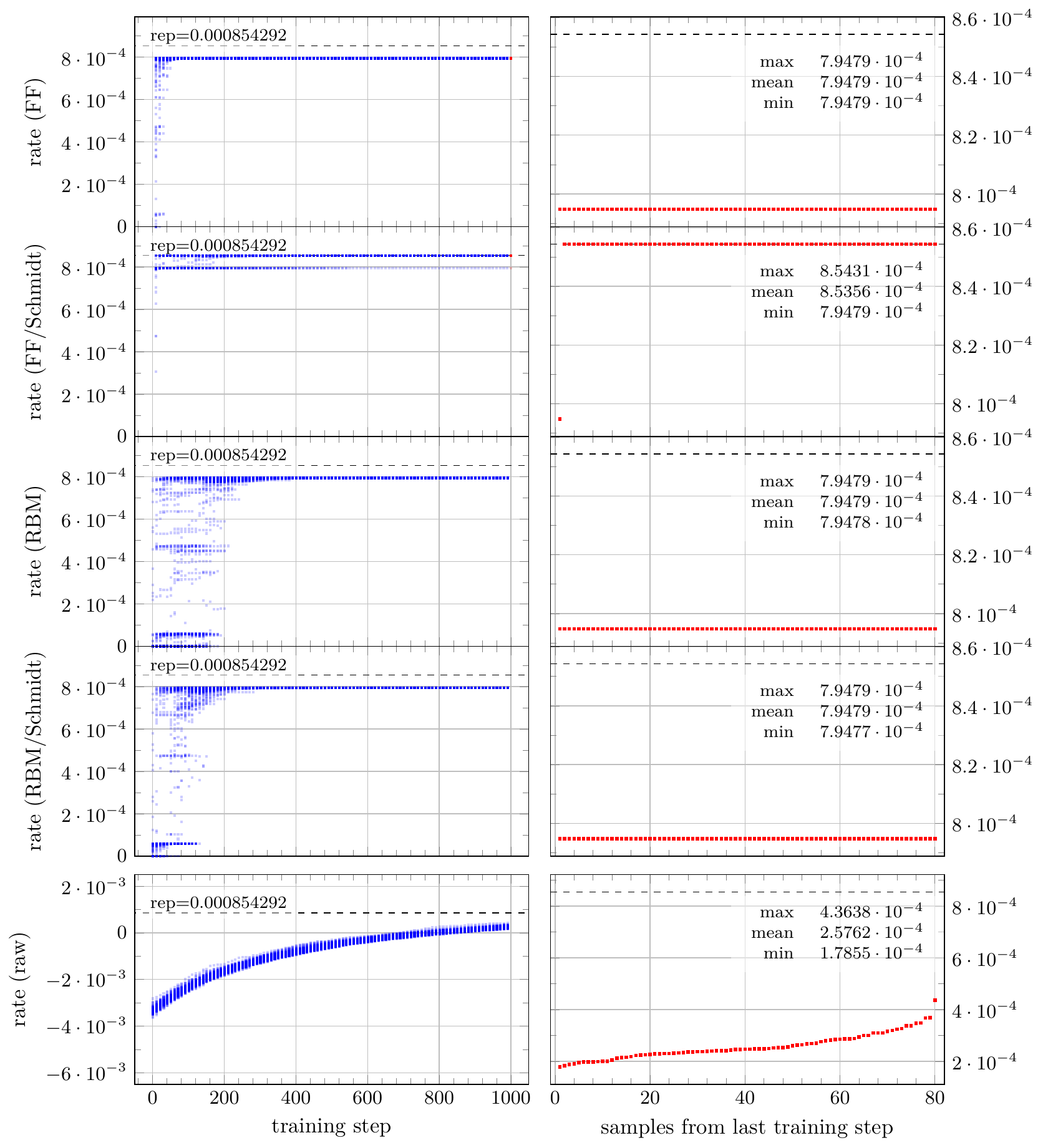}{an artificial bee colony (ABC) algorithm implemented in pagmo}{four}{We remark that ABC seems to have troubles moving beyond a local minimum around the three-repetition code with CI$=0.0007948$ in all but the FF/Schmidt ansatz.}
\extrafig{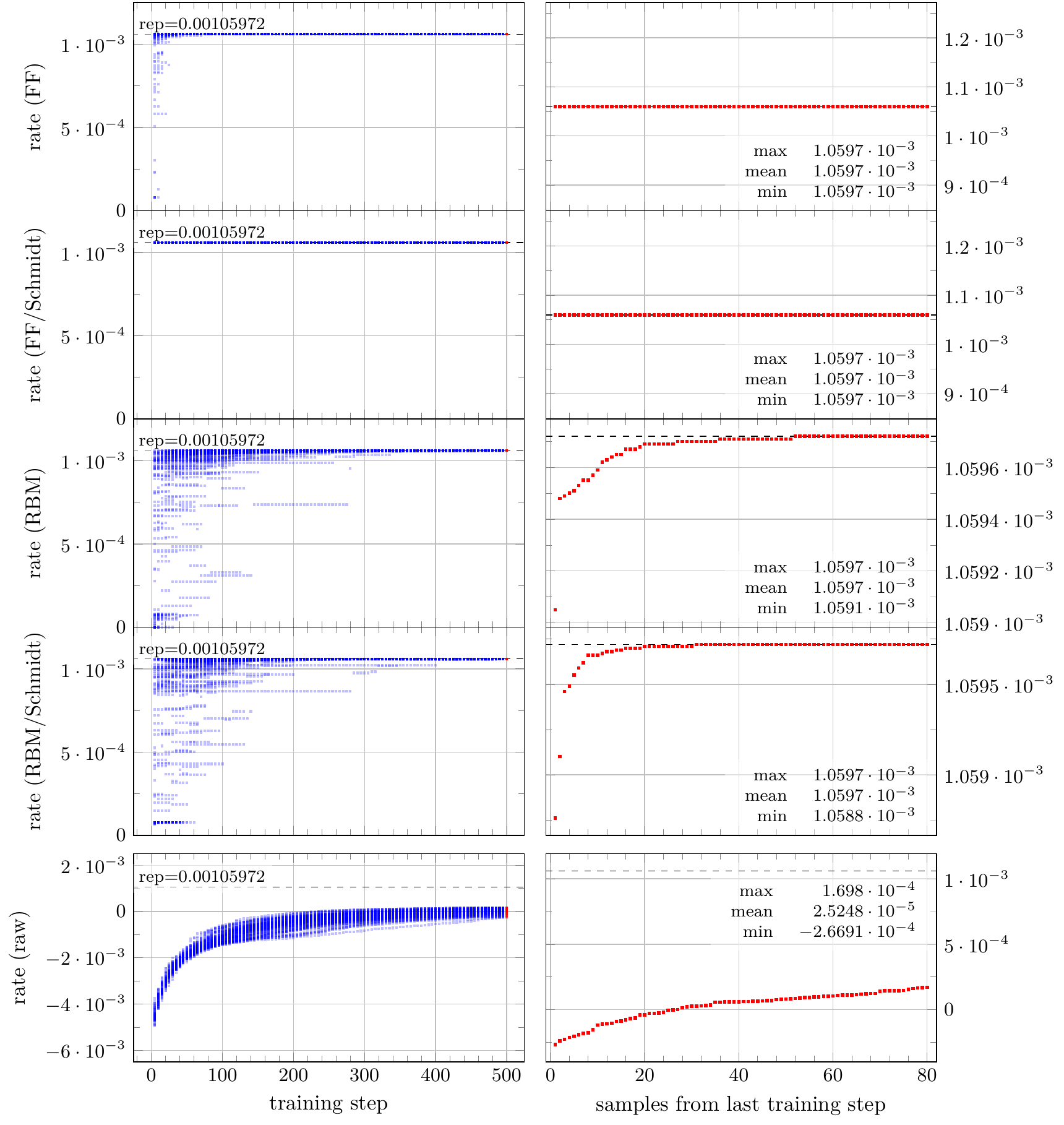}{a particle swarm (PSO) algorithm implemented in pagmo}{three}{}
\extrafig{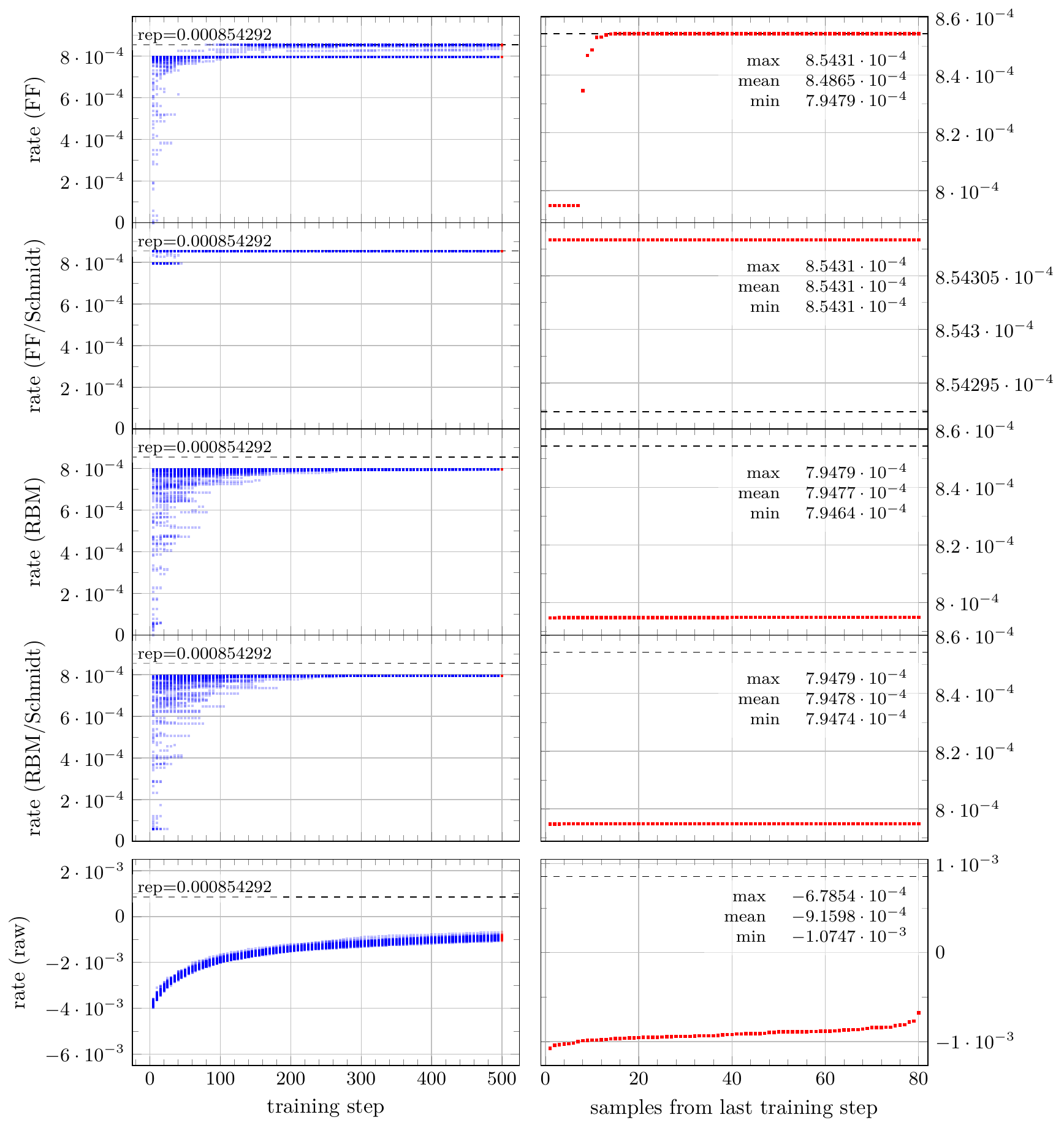}{a particle swarm (PSO) algorithm implemented in pagmo}{four}{}
\extrafig{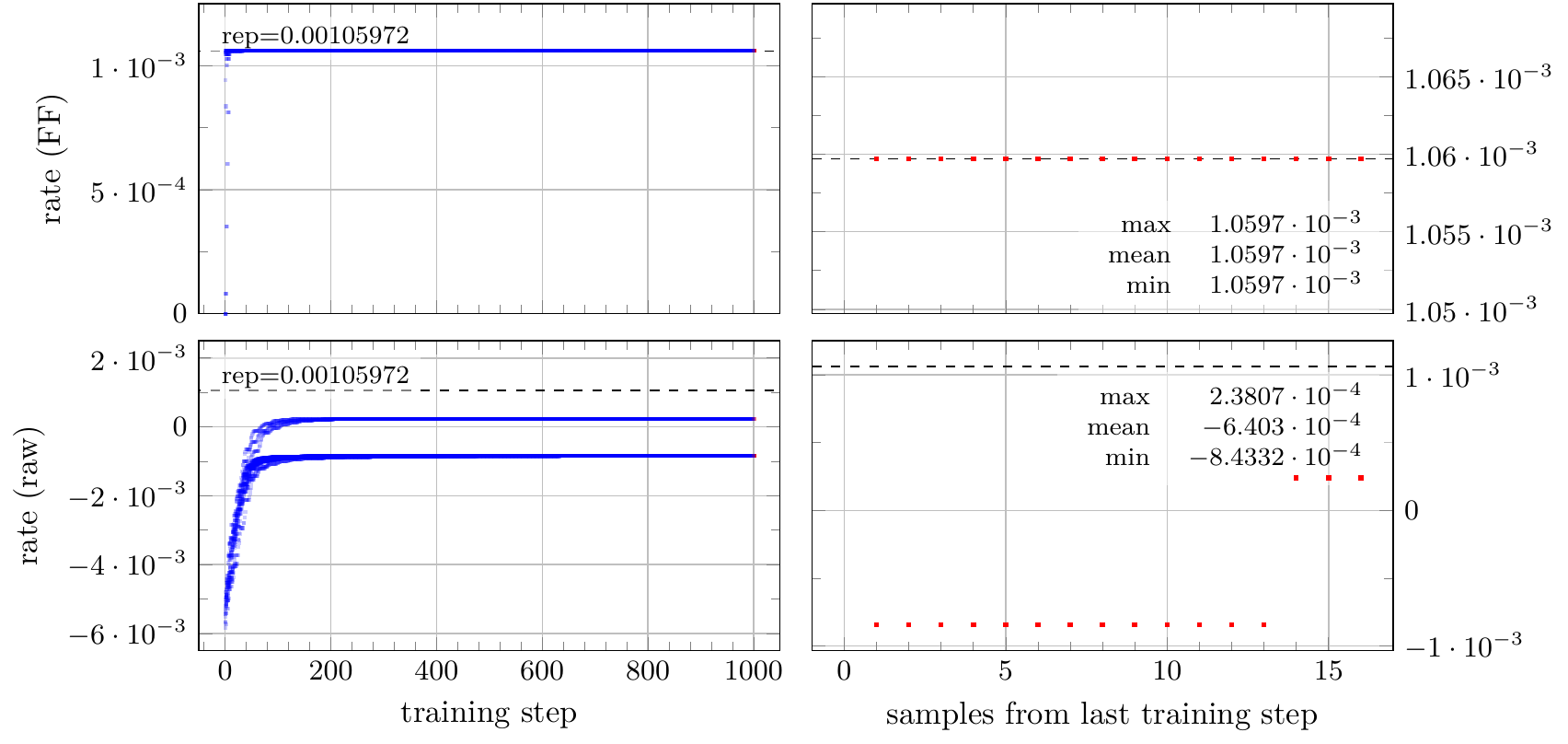}{a particle swarm (PSO) algorithm implemented in MATLAB}{three}{}
\extrafig{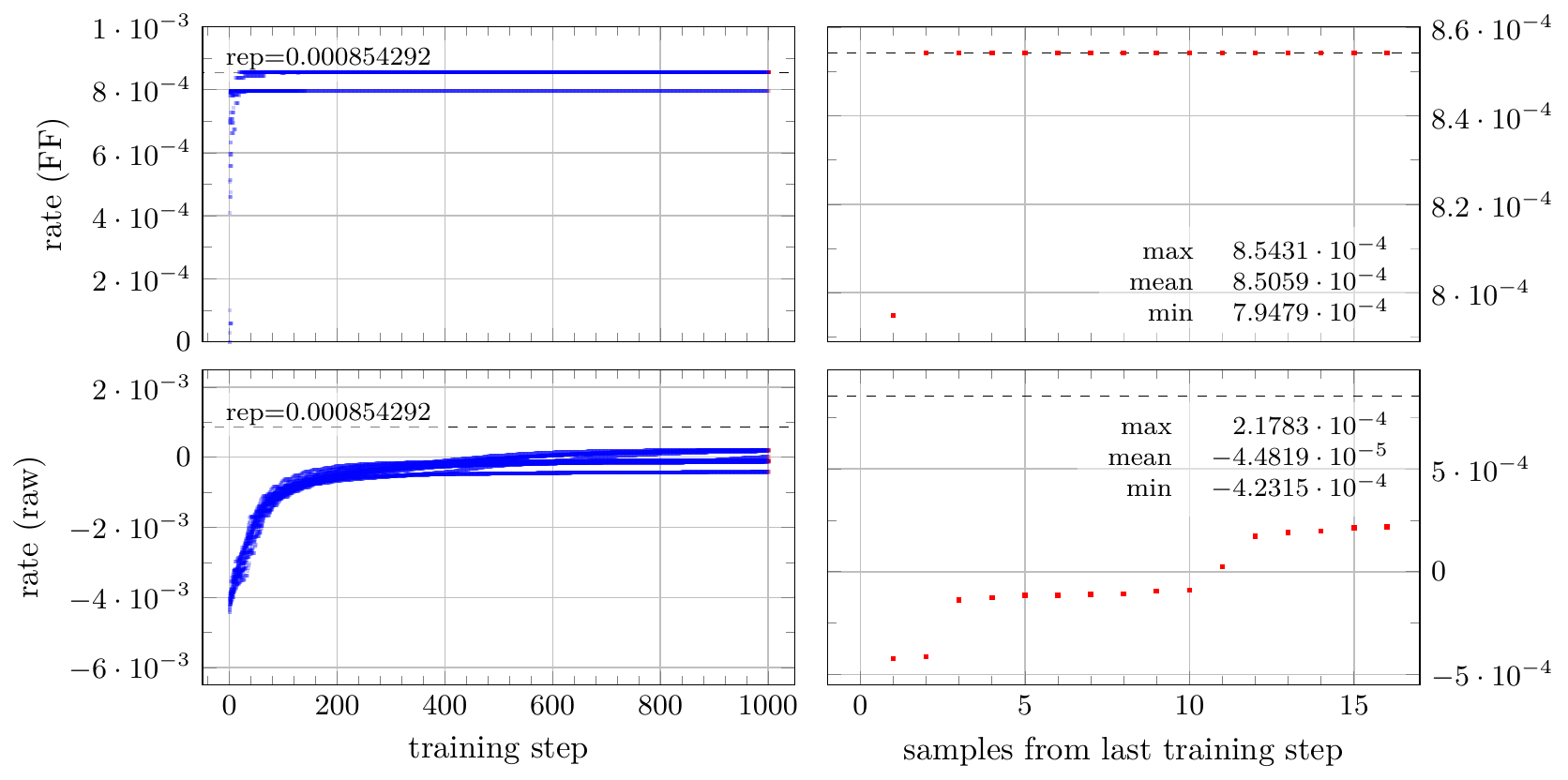}{a particle swarm (PSO) algorithm implemented in MATLAB}{four}{}
\extrafig{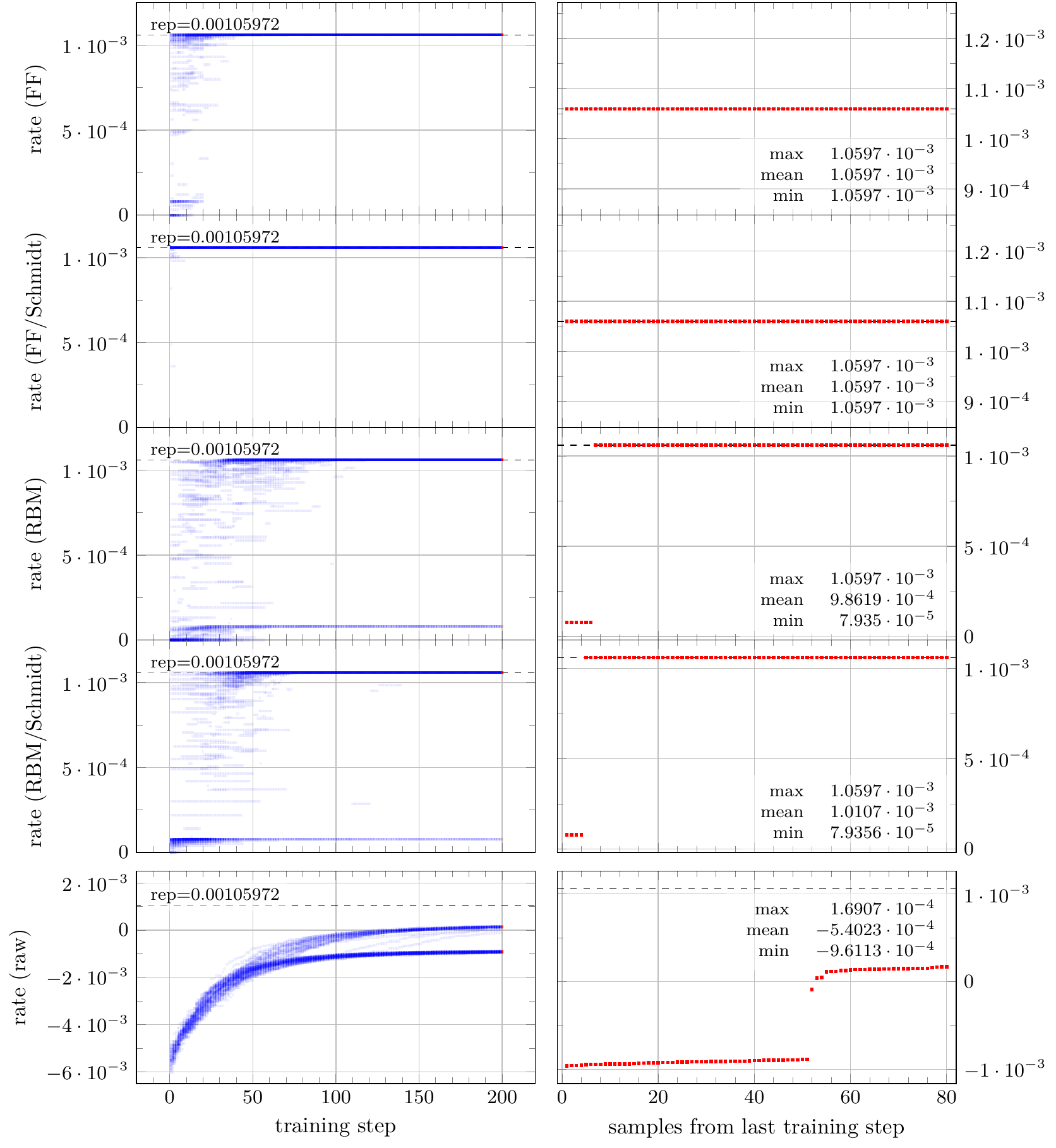}{a simple genetic (SGE) algorithm implemented in pagmo}{three}{}
\extrafig{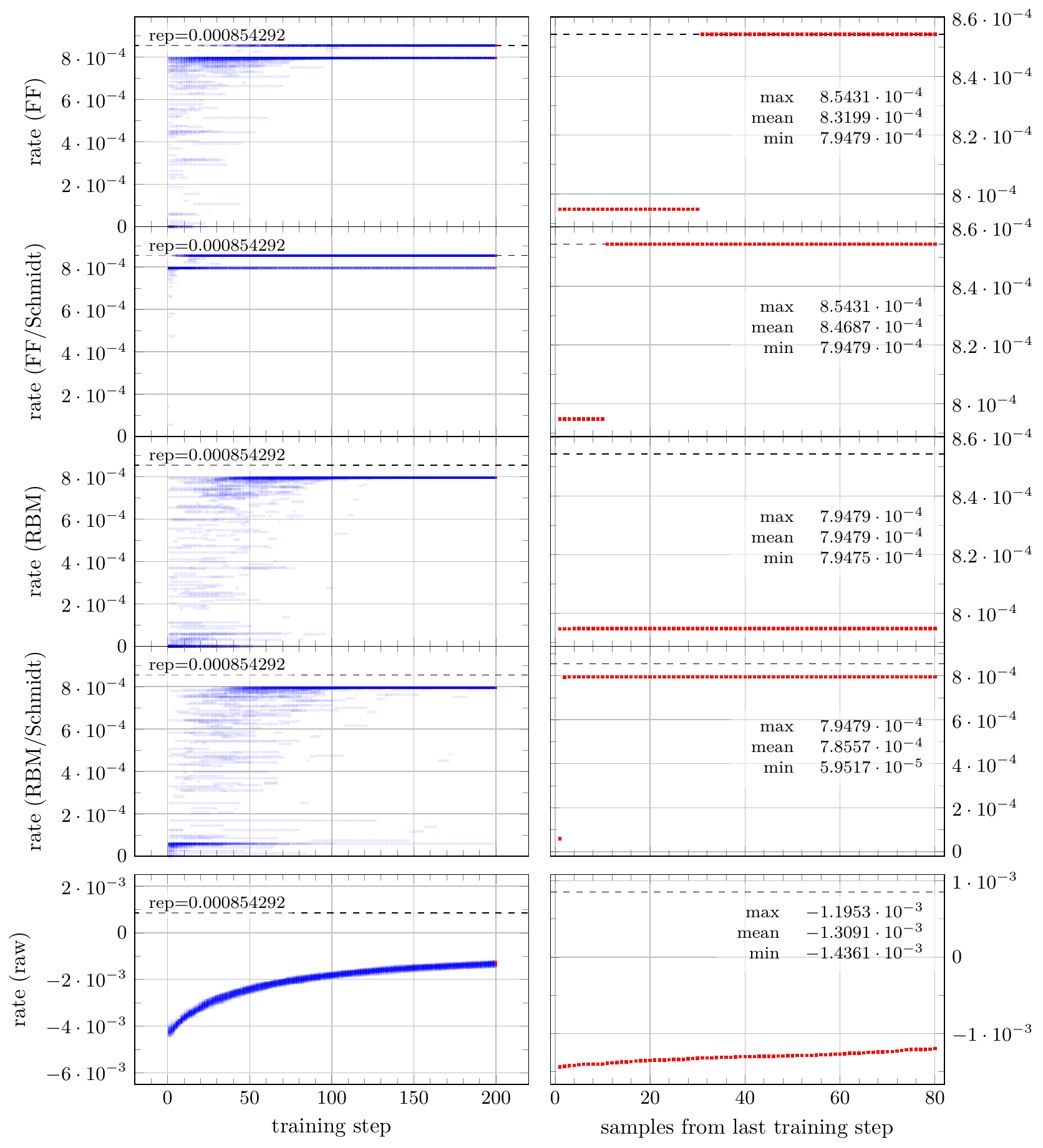}{a simple genetic (SGE) algorithm implemented in pagmo}{four}{}

\end{document}